\documentclass[letterpaper,twoside,12pt]{article}
\usepackage[dvips]{graphicx}
\usepackage{amsmath}
\usepackage{amssymb}
\usepackage[T1]{fontenc}
\usepackage{natbib}
\usepackage{authblk}
\newcommand{\uj}{\mathrm{j}}

\newcommand{\Real}{\mathrm{Re}}
\newcommand{\Imag}{\mathrm{Im}}

\title{Van Vleck correction generalization for complex correlators with multilevel quantization}

\author[1]{L. V. Benkevitch}
\author[1]{A. E. E. Rogers}
\author[1]{C. J. Lonsdale}
\author[1]{R. J. Cappallo}
\author[2]{D. Oberoi}
\author[1]{P. J. Erickson}
\author[3]{K. A. V. Baker}

\affil[1]{\small MIT Haystack observatory, Westford, MA 01886, USA.}
\affil[2]{\small National Centre for Radio Astrophysics, Ganeshkhind, Pune 411 007, Maharashtra, INDIA.}
\affil[3]{\small Carleton College, Northfield, MN 55057, USA}

\begin{document}

\maketitle

\begin{abstract}
Remote sensing with phased antenna arrays is based on measurement of the cross-correlations between the signals from each antenna pair. The digital correlator response to quantized inputs has systematic errors due to the information loss in the process of quantization. The correlation errors allow substantial abatement based on the assumption that the analog signals are stochastic processes sampled from a statistical distribution (usually the Gaussian). The correlation correction technique is named after Van Vleck who was the first to apply it to two-level clipping quantizers. The correction is especially important for high correlation levels, e.g. in studies of solar radio emissions. We offer a generalized method that for every antenna pair inputs the quantized signals' covariance and standard deviations, and outputs high-precision estimates of the analog correlation. Although correlation correction methods have been extensively investigated in the past, there are several problems that, as far as we know, have not been published yet, and that we present solutions to here. We consider a very general quantization scheme with arbitrary set of transition thresholds and output levels, and our correction method is designed for correlations obtained from signals with generally unequal standard deviations. We also provide a method for estimation of the analog standard deviation from the quantized one for subsequent use in the correlation correction. We apply the correction to the the complex-valued analytic signals, overwhelmingly used in modern remote sensing systems with arrays of antennas. The approach is valid not only for analytic signals with the imaginary part being the Hilbert transform of the real one, but also for more general, circularly symmetric complex processes whose real and imaginary parts may have arbitrary relationships to each other. This work was motivated by the need for greater precision in analysis of data from the Murchison Widefield Array (MWA).

\end{abstract}

\tableofcontents

\section{Introduction}

This study was motivated by ongoing efforts in the use of the Murchison Widefield Array (MWA) for solar and heliospheric science.  The MWA is a radio interferometer array in Western Australia featuring a number of innovations. Designed to cover the low radio frequency range of 80--300 MHz, it is intended to be instrumental in a variety of astronomy and astrophysics projects, such as the studies in the 21 cm neutral hydrogen line during the epoch of reionization (EoR); imaging of the sun and the inner heliospheric phenomena; study of transient radio sources; and study of a wide variety of discrete radio sources. The overall MWA design is described in the papers by  \citet{Lonsdale2009} and \citet{Tingay2013}. The MWA digital receiver is described by \citet{Prabu2015}, its correlator by \citet{Wayth2009} and \citet{Ord2015}, and its voltage capture system by \citet{Tremblay2015}. The latter also demonstrate the MWA capabilities in such areas as pulsar and solar science on sub-second time scales. The scientific applications of the MWA are detailed in the work by \citet{Bowman2013}. The MWA applications to solar and heliospheric science are discussed by \citet{OberoiBenkevitch2010,Oberoi2013,Oberoi2014}.

The MWA is designed to support high time and frequency resolution and excellent imaging dynamic range. In particular, the imaging dynamic range potential of the MWA for solar work is expected to be very high due to excellent instantaneous monochromatic 2-D coverage of spatial frequencies (the $uv$ plane), and dominance by a single localized region of emission (the solar disk and corona, <1 degree across), facilitating accurate self-calibration. This, however, has not yet been achieved. For example, \citet{Oberoi2013} show a set of solar images obtained on MWA, with dynamic ranges of typically 1000-1500, at least an order of magnitude lower than desired. This work deals with a leading candidate for the loss of dynamic range, namely errors in the cross-correlation computation due to quantization effects, and presents methods for their compensation. 

Interferometric radio imaging is based on computation of the complex visibilities at many points on the $uv$ plane. The visibility at each point is provided by a pair of antennas in the array, and it is proportional to the cross-correlation of the pair of signals. The visibilities are complex values, and require correlation of the complex-valued signals. Here we focus on real-valued correlations because the complex correlation can be obtained as two real correlations. The analog signals from the antennas are quantized at the Nyquist (or higher) rate and channelized by frequency using digital techniques, yielding streams of binary numbers. The digital correlator is a device or a piece of software that ingests these digital data streams and outputs their cross-products for every pair of the antennas in each of the frequency channels. These cross-products are averaged over an interval known as the accumulation period, yielding the covariances of the signal pairs. The correlations are computed as the covariances divided by the product of the RMS of the two signals. The quantized correlations obtained in this way are subject to digital information loss, and differ from the true correlations between the analog signals. 

\begin{figure}[ht] 
\noindent\includegraphics[width=20pc]{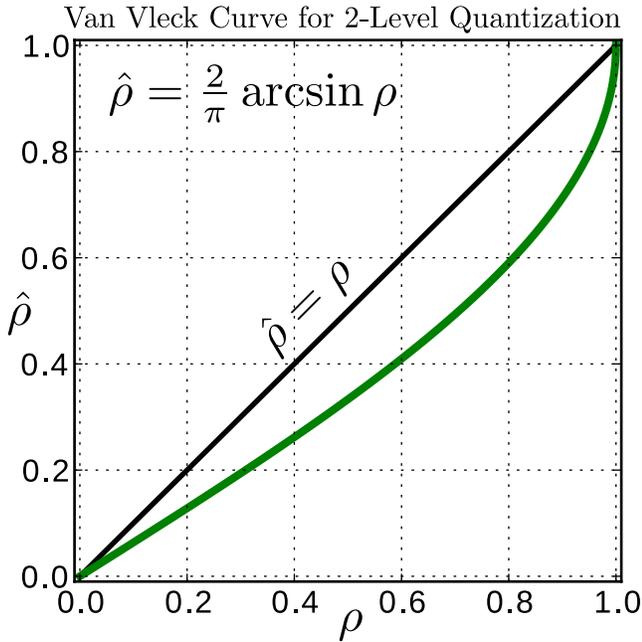}
\caption{\small Classical Van Vleck dependence of the correlation computed as the averaged product (covariance) of two signals clipped at two levels, -1 and +1, on the true, analog correlation $\rho$.}
\label{fig_classical_vanvleck_inverted}
\end{figure}
The classical \citet{Vanvleck1966} function in Fig.~\ref{fig_classical_vanvleck_inverted} shows the dependence of the quantized correlation, $\hat{\rho}$, on the analog correlation, $\rho$. Usually, this curve is plotted inverted, as $\rho = f(\hat{\rho})$, but we are interested in showing how the digital correlator transforms the analog correlation. The quantization scheme considered by Van Vleck is the simplest of all possible cases (see Fig.~\ref{fig_simple_qpatterns}, a): a simple signal clipping between the two levels, -1 and +1. The quantization patterns with more levels can provide less distortion in $\hat{\rho}$. However, the quantization distortion effect on the correlation in principle remain similar, although here it is more salient. One can see the quantized correlation is systematically lower than the analog correlation, $\hat{\rho}<\rho$. The curve slope differs from the ideal  $\rho = \hat{\rho}$. For small analog correlations it leads to a reduction in signal to noise ratio, and so the quantization impairs the sensitivity. Note that the slope of the curve changes sharply for analog correlations close to unity. In solar observations the correlations are high, and distortions in the correlation near unity can result in large departures from linearity. This non-linearity causes different levels of digital error on different baselines, in a manner that in general cannot be decomposed into antenna-based complex gains.  Radio interferometric imaging and precision calibration packages normally assume that errors are almost entirely antenna-based.  The violation of that assumption in the case of strong correlations and significant digital information losses, as described here, is a major contributor to limitations in achievable imaging dynamic range.

The \citet{Vanvleck1966} formula $\rho = \sin((\pi/2)\hat{\rho})$ can be used for the quantized correlation correction, but only for 2-level quantizers. The Van Vleck correction for all other quantization patterns cannot be rendered in closed form. \citet{Cooper1970} was one of the first who considered 2-bit correlators with the quantization pattern given in Fig.~\ref{fig_simple_qpatterns}~(b). His paper contains derivations and plots of the 2-bit correlator sensitivities relative to analog correlation as functions of the switching threshold $v_0$, for different values of the level $n$. Also, he provided graphs of the dependencies $\rho = f(\hat{\rho})$ for different $v_0$ and $n$ -- yielding curves similar to that of Van Vleck. \citet{Hagen1973} analyzed several different quantization schemes and provided a table of the formulas relating the correlator output and the analog correlations for eleven cases. The questions of correct correlation estimation for a 3-level digital correlator were explored by \citet{Kulkarni1980}. They provided exact formulas relating the correlator output and the analog signal correlation and described approximation schemes for fast calculations in real time. The efficiency of a digital correlator is defined as the RMS of the quantized correlation relative to the RMS of an ideal analog correlation, and \citet{Thomson2007} provide formulas for the quantization efficiency which depends on the choice of quantization levels and switching thresholds. The work by \citet{Johnson2013} is devoted to design strategies for the correlators to achieve optimal efficiency. A comprehensive discussion of digital correlators, their efficiency, and the approximation formulas can be found in the book by \citet[Chapter 8]{TMS2001}.

The work described in this paper is based on the aforementioned literature. It builds on previous work in three ways. First, we make our derivations assuming the correlator inputs have generally different RMS (or STD). This is important for MWA as it makes the Van Vleck correction substantially independent of the antenna gains settings. In earlier publications the correlator inputs have been assumed to have equal RMS. Second, unlike our predecessors, we derive a fully general formula relating the correlator output (the covariance) and the true, analog correlation, which accommodates an arbitrary, generally irregular quantization scheme. Third, we emphasize the fact that the standard deviations computed using the quantized signals have significant offset with respect to the STD of the analog signal. Therefore, the analog STDs, required for the Van Vleck correction, must be estimated using the quantized STDs. We derive convenient formulas for such estimations.  

The outline of this paper is as follows. Section~2 provides the basis for the subsequent derivations. It gives the formulas and the terminology to be used throughout the paper. Section~3 introduces the quantization patterns (or characteristics, or schemes) to be used further. In Section~4 we derive the main formula for the Van Vleck correction (see Eq.~\eqref{qkappa_as_integral}) and its simplifications for some specific cases. In Section~5 the formulas for STD estimation are derived. Section~6 explains the details of the software implementation of the Van Vleck correction for real and complex correlators.  It also provides description of the simulation  and results. In Section~7 we show that Eq.~\eqref{qkappa_as_integral}) can be easily transformed into the formulas for particular quantization schemes given in \citet[Chapter 8]{TMS2001}. We also discuss some interesting properties of the Van Vleck correction functions.

\section{Definitions and formulation}

The covariance $\kappa$ of two band-limited analog signals $x(t)$ and $y(t)$ for zero delay between them is their product averaged over some period of time:
\begin{equation}
  \label{covariance}
  \kappa = \langle x(t)y(t) \rangle.
\end{equation}
The covariances of each signal with itself are called variances,
\begin{equation}
  \label{variances}
  \sigma^2_x = \langle x(t)^2 \rangle, \quad \sigma^2_y = \langle y(t)^2 \rangle,
\end{equation}
and their square roots $\sigma_x$ and $\sigma_y$ are named standard deviations (STD) or root mean squares (RMS):
\begin{equation} \label{stds}
  \sigma_x = \sqrt{\langle x(t)^2 \rangle}, \quad \sigma_y = \sqrt{\langle y(t)^2 \rangle}.
\end{equation}
The correlation coefficient of $x(t)$ and $y(t)$ is the result of their covariance normalization
\begin{equation} 
  \label{correl}
  \rho = \frac{\langle x(t)y(t) \rangle}{\sigma_x \sigma_y} = \frac{\kappa}{\sigma_x \sigma_y}.
\end{equation}
The analog-to-digital converter (ADC) turns the input analog signal $x$ into its quantized counterpart $\hat{x}$, which is a stream of binary numbers. To preserve the information transmitted within the frequency band the ADC must sample the input signal at the Nyquist rate equal to the doubled band width in the case of a purely real signal. For each pair of such data streams $\hat x$ and $\hat y$ the correlator fulfills the repeated operation ``multiply-add" over the specified number of samples $N$, corresponding to the ``integration time", forming the raw correlator output
\begin{equation}
  N\hat{\kappa} = \sum\limits_i \hat{x}_i\hat{y}_i.
\end{equation}
For radio astronomical applications, the arrays of these values representing the results of the observations are typically combined with extensive header information and stored in {\tt uvfits} files \citep{Greisen2003, Sault1995} or in CASA measurement set databases \citep{McMullin2007}. 
The averaging of $N\hat{\kappa}$ renders what we will call ``the quantized covariance'' $\hat{\kappa}$:
\begin{equation}
  \label{qcovariance}
  \hat{\kappa} = \langle \hat{x}\hat{y} \rangle = \frac{1}{N} \sum\limits_i \hat{x}_i\hat{y}_i.
\end{equation}
Typically, the correlators compute the autocovariances too, which we will call ``quantized variances'' and use for obtaining the ``quantized STDs":
\begin{equation}
  \label{qstd}
  \hat{\sigma}_x = \sqrt{\langle \hat{x}^2 \rangle}, \quad \hat{\sigma}_y = \sqrt{\langle \hat{y}^2 \rangle}.
\end{equation}
The quantized correlation $\hat \rho$ is calculated similarly to the analog correlation $\rho$:
\begin{equation} 
  \label{qcorrel}
  \hat{\rho} = \frac{\langle \hat{x}\hat{y} \rangle}{\hat{\sigma}_x \hat{\sigma}_y} = 
      \frac{\hat{\kappa}}{\hat{\sigma}_x \hat{\sigma}_y}.
\end{equation}
The loss of information due to the quantization leads to systematic errors in $\hat{\kappa}$ and $\hat{\rho}$. However, these losses and associated errors can be compensated for. With good precision each pair of the antenna signals $x(t)$ and $y(t)$ can be treated as a two-dimensional zero-mean Gaussian stochastic process with values $(x,y)$ sampled from the joint normal probability distribution function (PDF) $f(x,y,\sigma,\rho)$:
\begin{equation} 
  \label{normal2d_pdf}
  f(x,y,\sigma,\rho) = \frac{1}{2\pi\sigma_x\sigma_y\sqrt{1-\rho^2}} \exp\left\{-\frac{1}{2(1-\rho^2)} 
        \left[ \frac{x^2}{\sigma_x^2} + \frac{y^2}{\sigma_y^2} - \frac{2\rho x y}
                                         {\sigma_x\sigma_y} \right] \right\},
\end{equation}
where $\sigma_x$ and $\sigma_y$ are the $x$ and $y$ STDs as in \eqref{stds}, and $\rho$ is the correlation as in \eqref{correl}. The information contained in the PDF can be used for estimation of the true correlation coefficient $\rho$.

\section{Quantization patterns}

\begin{figure}[ht] 
\noindent\includegraphics[width=20pc]{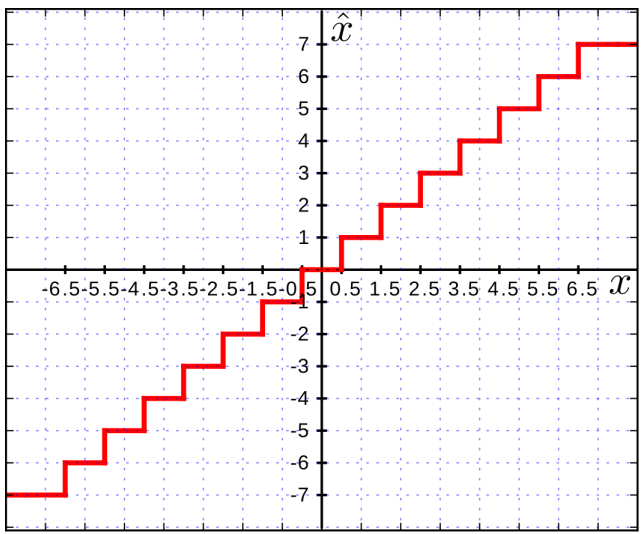}
\caption{\small Quantization characteristics of a 4-bit analog-to-digital converter (ADC) with an odd number, $2^4-1=15$, of levels. A quantizer of this type is assumed as the pre-correlator stage.}
\label{adc_4bit_characteristics}
\end{figure}

Modern remote sensors, such as radars, radio interferometers or radio telescopes, tend to use analog/digital converters (ADCs) with a relatively large number of quantization levels, and equal steps between the levels leading to what is known as a regular quantization pattern. Its output is usually characterized not by the number of quantization levels, but by its precision -- the number of bits in the output word. With the equally spaced levels the output numbers are proportional to the input voltages. Regular quantization can be of two kinds: with even or odd numbers of levels \citep{Thomson2007}. Quantizations with an even number of levels have the transition thresholds at the consecutive integer numbers including zero, while for an odd number of levels the thresholds occur at the integers $\pm\tfrac{1}{2}$. The even number of levels does not include the zero level, whereas the odd number of levels includes the zero. As an example of the system with odd number of levels, the Murchison Widefield Array \citep{Ord2015, Prabu2015, Tingay2013, Lonsdale2009} at different stages of its data paths uses 8 bits (255 levels), 5 bits (31 levels), and 4 bits (15 levels). The N-bit signed binaries have the lower and upper values
\begin{equation} \label{maxlevel}
  \pm m = \pm (2^{N-1}-1),
\end{equation}
i.e. all the output values are within the interval $[-m .. m]$. In the MWA case we have the 4-bit integers $N = 4$, $m = 2^3-1=7$, so the 4-bit integers can have values within [-7 .. 7]. The ADC quantizes its input analog signal, $x(t)$ into a stream of signed integers, $\hat{x}_i$, according to the characteristic curve (or ``staircase") shown in Fig.~\ref{adc_4bit_characteristics}. In the odd-level quantization scheme used in MWA the transitions between discrete output levels occur at the half-integer boundaries of the input signal. For example, if the input is within the interval [1.5 .. 2.5), the output is 2. If the input drops below 1.5, the ADC output switches to 1. If the input rises above 2.5, the ADC output switches to 3. In case of a complex analytic signal  $z(t) = x(t) + \mathrm{j} y(t)$, its quantized counterpart is denoted as $\hat{z} = \hat{x} + \mathrm{j} \hat{y}$. The quantized complex signals are represented as pairs of 4-bit real and 4-bit imaginary components. They both obey the same rule: $\hat{x}, \hat{y} \in$ [-7 .. +7]. 

\begin{figure}[ht] 
\noindent\includegraphics[width=20pc]{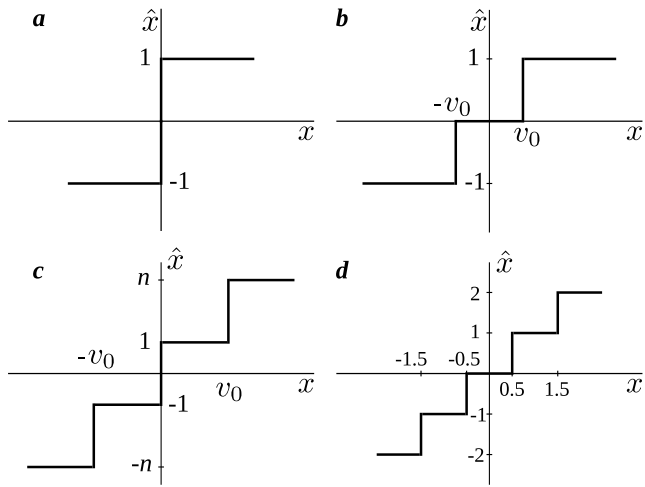}
\caption{\small Some simple quantization patterns for (a) two-level (clipping) quantizer, (b) three-level, (c) four-level, and (d) five-level quantizers.}
\label{fig_simple_qpatterns}
\end{figure}

\begin{figure}[ht] 
\noindent\includegraphics[width=22pc]{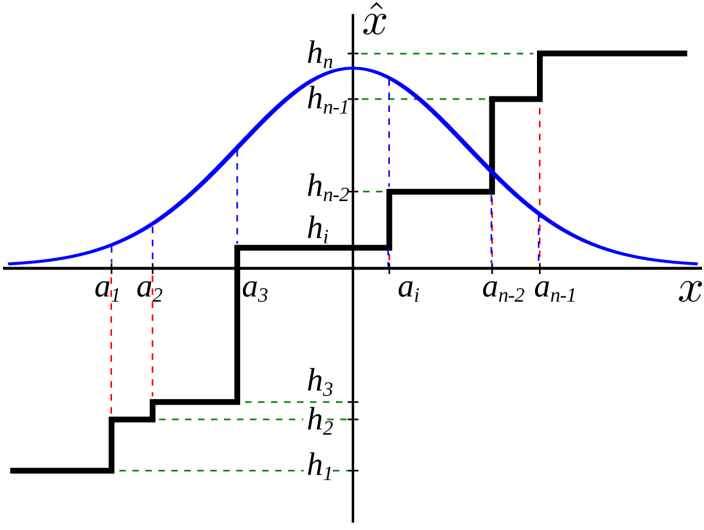}
\caption{\small General quantization pattern $adc = \{n,a,h\}$, with $n$ levels of quantization, $h = h_1, h_2, \ldots, h_n$, and $n-1$ switching thresholds, $a = a_1, a_2, \ldots, a_{n-1}$. The Gaussian PDF is shown as the bell-shaped curve.}
\label{gen_qpattern}
\end{figure}

In past decades quantization patterns with fewer number of sometimes non-equidistant level were commonly used. Some of them are shown in Fig.~\ref{fig_simple_qpatterns}. These patterns are still used, typically in very high speed systems with large number of data channels to reduce the amount of data to be processed in real time. With a small number of levels the spacing between them can be optimized. Although we focus here on the MWA 4-bit quantizer, we do not restrict our study to the regular patterns only. Instead, we consider a quantizer with $n$ levels $h_i = h_1, h_2, \ldots, h_{n}$ and $n-1$ switching thresholds $a_i = a_1, a_2, \ldots, a_{n-1}$. We will call this a general (or arbitrary) quantization pattern and denote it as 
\begin{equation} 
  \label{qpattern}
  \mathrm{adc} = \{n, a, h \}.
\end{equation}
An example of a general pattern is shown in Fig.~\ref{gen_qpattern}.

\section{Digital correlator response as a function of analog correlation}

Here ``correlator'' means ``correlator of real signals''. The relationship derived in this Section will be used later to make the Van Vleck correction of the analytic (complex) signals, used in modern radio interferometers.

Following \citet{Hagen1973} we base our derivations on Price's theorem \citep{Price1958}. Since its initial publication the theorem has been generalized and its proof simplified \citep{Brown1967,Papoulis1965,McMahon1964}. In one of its generalizations the theorem is written as follows:
\begin{equation} 
  \label{gen_price}
  \frac{\mathrm d \langle g \rangle}{\mathrm d \kappa} = \left\langle \frac{\partial^2 g}{\partial x \partial y} \right\rangle,
\end{equation}
where $g = g(x,y)$ is an arbitrary non-linear function of correlated random variables $x$ and $y$, jointly distributed according to the Gaussian PDF~\eqref{normal2d_pdf}, and $\kappa$ is their covariance~\eqref{covariance}. Substitution $g(x,y) = \hat{x}\hat{y}$, $\kappa = \sigma_x\sigma_y\rho$, and the unnormalized correlator output $\hat{\kappa}$ from~\eqref{qcovariance} yields 
\begin{equation} 
  \label{price}
  \frac{\mathrm d \hat{\kappa}}{\mathrm d \rho} = 
    \sigma_x\sigma_y\left\langle \frac{\partial\hat{x}}{\partial x} \frac{\partial \hat{y}}{\partial y} \right\rangle.
\end{equation}
To expand the right-hand side, one can notice that the derivatives $\partial\hat{x}/\partial x$ and $\partial\hat{y}/\partial y$ are expressed as weighted sums of the delta-functions 
\begin{equation} 
  \label{derivs_xy}
  \frac{\partial\hat{x}}{\partial x} = \Delta h_1\delta(x-a_1) + \Delta h_2\delta(x-a_2) + \ldots 
                                     + \Delta h_{n-1}\delta(x-a_{n-1})
\end{equation}
where $\Delta h_i = h_{i+1} - h_i$ are the level distances. The formula for $\partial\hat{y}/\partial y$ is similar. Their product is
\begin{equation} 
  \label{deriv_prod}
  \frac{\partial\hat{x}}{\partial x} \frac{\partial\hat{y}}{\partial y} = 
      \sum\limits_{i=1}^{n-1} \sum\limits_{k=1}^{n-1} \Delta h_i \Delta h_k \delta(x-a_i) \delta(y-a_k).
\end{equation}
The right-hand side of Eq.~\eqref{price} is the expectation of this product of derivatives. The expectation of each $ik$-th term is equal to its magnitude $\Delta h_i \Delta h_k$ multiplied by its probability of occurrence $P(a_i,a_k)$:
\begin{equation} 
  \label{expect_of_deriv_prod}
  \left\langle \frac{\partial\hat{x}}{\partial x} \frac{\partial\hat{y}}{\partial y} \right\rangle = 
      \sum\limits_{i=1}^{n-1} \sum\limits_{k=1}^{n-1} \Delta h_i \Delta h_k \, P(a_i,a_k).
\end{equation}
This probability is expressed as the double integral over the whole $xy$ plane of the delta-functions multiplied by the normal distribution density function ~\eqref{normal2d_pdf}:
\begin{equation} 
  \label{int_delta}
  P(a_i,a_k) = \iint\limits_{-\infty-\infty}^{+\infty+\infty} \hspace{-0.1cm}
               \delta(x-a_i) \delta(y-a_k) f(x,y) \, \mathrm{d}x\mathrm{d}y.
\end{equation}
Due to the delta function properties, such integrals are easily found as the values of the PDF at the positions of the integrand's delta functions:
\begin{equation} 
  \label{int_delta_solved}
  P(a_i,a_k) = f(a_i,a_k).
\end{equation}
In more detail, probabilities of the right-hand side terms of the Price's theorem \eqref{price} are
\begin{equation} 
  \label{term_prob}
   P(a_i,a_k) = \frac{1}{2\pi\sigma_x\sigma_y\sqrt{1-\rho^2}} \exp\left\{-\frac{1}{2(1-\rho^2)} 
      \left[ \frac{a_i^2}{\sigma_x^2} + \frac{a_k^2}{\sigma_y^2} - \frac{2\rho a_i a_k}{\sigma_x\sigma_y}   \right]
      \right\}.
\end{equation}
Substituting the latter into \eqref{price} and solving for the quantized covariance $\hat{\kappa} = \langle \hat{x}\hat{y} \rangle$ gives us the expression for the Van Vleck correction function for correlators with general quantization patterns \eqref{qpattern}, shown in Fig.~\ref{gen_qpattern}, as
\begin{equation} 
  \label{qkappa_as_integral}
   \hat{\kappa} = \frac{1}{2\pi} \int\limits_0^\rho
      \frac{1}{\sqrt{1-\zeta^2}}  \sum\limits_{i=1}^{n-1} \sum\limits_{k=1}^{n-1} \Delta h_i \Delta h_k  
      \exp\left\{-\frac{1}{2(1-\zeta^2)} 
      \left[ \frac{a_i^2}{\sigma_x^2} + \frac{a_k^2}{\sigma_y^2} - \frac{2\zeta a_i a_k}{\sigma_x\sigma_y}   \right]
      \right\}  \mathrm{d}\zeta,
\end{equation}
where $n$ is the number of quantization levels. Numerical integration of the latter lets us predict the normalized digital correlator output $\hat{\kappa}$ for any known correlation $\rho$ and standard deviations $\sigma_x$ and $\sigma_y$ of two analog signals $x(t)$ and $y(t)$. 

For an N-bit correlator with the regular quantization pattern as that of MWA in Fig.~\ref{adc_4bit_characteristics} the Van Vleck correction relationship takes this more concrete form:
\begin{align} 
  \label{qkappa_as_integral_mwa}
   \hat{\kappa} = \frac{1}{2\pi} \int\limits_0^\rho
      \frac{1}{\sqrt{1-\zeta^2}}  \sum\limits_{i=-m}^{m-1} \sum\limits_{k=-m}^{m-1} \exp&\left\{-\frac{1}{2(1-\zeta^2)} 
      \left[ \frac{(i+\tfrac{1}{2})^2}{\sigma_x^2} + \frac{(k+\tfrac{1}{2})^2}{\sigma_y^2}  \right. \right. \nonumber \\
      &- \left. \left. \frac{2\zeta(i+\tfrac{1}{2})(k+\tfrac{1}{2}) }{\sigma_x\sigma_y}   \right]
      \right\}  \mathrm{d}\zeta,
\end{align}
where $m$ is the maximum number of the positive quantization levels for an N-bit ADC, found by Eq.~\eqref{maxlevel}. For the MWA case, $m=7$. 

Eqs.~\eqref{qkappa_as_integral} and~\eqref{qkappa_as_integral_mwa} express the functional dependence 
\begin{equation} 
  \label{qkappa_as_fun_rho}
  \hat{\kappa} = g(\rho,\sigma_x,\sigma_y).
\end{equation}
Conversely, we need to determine the true analog correlation from the correlator output using the inverse of  \eqref{qkappa_as_integral}
\begin{equation} 
  \label{rho_as_fun_qkappa}
  \rho = g^{-1}(\hat{\kappa},\sigma_x,\sigma_y).
\end{equation}
The inverse function $g^{-1}(\cdot)$ can be built in the form of a three-dimensional table with the dimensions $\hat{\kappa}$, $\sigma_x$, and $\sigma_y$, the cells of which contain the values of analog correlation $\rho$.

\section{Estimation of the standard deviations of analog signals}

The STDs of the quantized signals are not equal to those of the analog signals. Estimation of the true analog correlation~\eqref{correl} (called the Van Vleck correction), based on the joint normal distribution \eqref{normal2d_pdf}, requires knowledge of the analog standard deviations of the signals, $\sigma_x$ and $\sigma_y$. Their quantized counterparts, $\hat{\sigma}_x$ and $\hat{\sigma}_y$, obtained by formulas~\eqref{qstd} from autocovariances found in the correlator output, contain systematic quantization errors. Therefore, the analog STDs need to be estimated themselves before being used in the correlation estimation. Below we provide a method of the STD estimation.

Let us express the quantized STD $\hat{\sigma}$  as a function of the true analog STD $\sigma$:
\begin{equation} 
  \label{qsig_as_fun_sig}
  \hat{\sigma} = f(\sigma).
\end{equation}
Consider an ADC with arbitrary quantization pattern \eqref{qpattern} (see Fig.~\ref{gen_qpattern}) and a quantized waveform $\hat{x} = \mathrm{adc}(x(t))$. The quantized variance $\hat{\sigma}^2$ is expectation of the correlator response to two identical inputs, $\hat{\sigma}^2 = \langle \hat{x}\hat{x} \rangle$. The analog signal $x(t)$ is a random process, and it is a valid approximation to assume that it is sampled from the zero-mean Gaussian distribution with the PDF
\begin{equation} 
  \label{normal_pdf}
  \phi(x,\sigma) = \frac{1}{\sigma\sqrt{2\pi}} \exp\left(-\frac{x^2}{2\sigma^2}\right).
\end{equation}
The probability that $x(t)\leqslant a$ is evaluated through the normal cumulative distribution function (CDF) as
\begin{equation} 
  \label{normal_cdf}
  \Phi\left(\frac{a}{\sigma}\right) = \frac{1}{\sigma\sqrt{2\pi}} 
      \int\limits_{-\infty}^{a} \exp\left(-\frac{t^2}{2\sigma^2}\right) \mathrm{d}t =
      \frac{1}{2}\left[1 + \mathrm{erf}\left(\frac{a}{\sigma\sqrt{2}}\right) \right],
\end{equation}
The expectation of the correlator output is therefore the sum of products of the probability that $x$ is within an interval between the switching thresholds, $[a_i,a_{i+1}]$, and the output value in this interval, $h^2_i$. As illustrated in Fig.~\ref{gen_qpattern}, within each interval the probability $P\left(x \in [a_i,a_{i+1}]\right)$ is equal to the area between the $x$ axis and the normal PDF curve \eqref{normal_pdf}, bounded by the vertical walls. This probability can be expressed using the CDF \eqref{normal_cdf}:
\begin{equation} 
  \label{prob_interv}
  P\left(x \in [a_i,a_{i+1}]\right) = \Phi\left(\frac{a_{i+1}}{\sigma}\right) - \Phi\left(\frac{a_i}{\sigma}\right).
\end{equation}
With the open intervals $(-\infty,a_1]$ and $[a_{n-1},\infty)$, we get the desired function \eqref{qsig_as_fun_sig}: 
\begin{equation} 
  \label{qsig_general}
  \hat{\sigma} = \left( h_1^2\Phi\left(\frac{a_1}{\sigma}\right) + h_n^2\left[1-\Phi\left(\frac{a_{n-1}}{\sigma}\right)\right] 
     + \sum\limits_{i=2}^{n-1} h_i^2 \left[ \Phi\left(\frac{a_i}{\sigma}\right) -\Phi\left(\frac{a_{i-1}}{\sigma}\right) \right]    
     \right)^{1/2}.
\end{equation}
In order to estimate an analog STD this function needs to be inverted and tabulated (or, in simple cases, interpolated) as
\begin{equation} 
  \label{sig_as_fun_qsig}
  \sigma = f^{-1}(\hat{\sigma}).
\end{equation}

\begin{figure}[ht] 
\noindent\includegraphics[width=22pc]{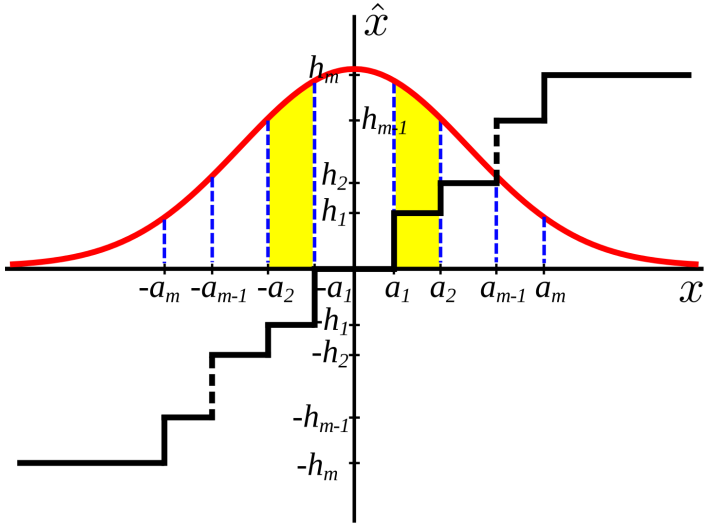}
\caption{\small  Normal probability density law over the the intervals of equal correlator output for an odd (centrosymmetric) quantization pattern. The probability that the correlator response is $h_1^2$ equals to the sum of areas of two highlighted curvilinear trapezoids over $(-a_2,-a_1]$ and $[a_1,a_2)$ intervals.}
\label{odd_erfs_levels}
\end{figure}

For a general odd (or centrosymmetric) quantization pattern shown in Fig.~\ref{odd_erfs_levels} the function~\eqref{qsig_general} can be somewhat simplified. The axisymmetrical (with respect to the $\hat{x}$ axis) area under the PDF curve over any $[-a,a]$ interval is the probability $P\left(x \in [-a,a] \right)$, which we denote as 
\begin{equation} 
  \label{prob_sym}
   \Psi(a)= \Phi\left(\frac{a}{\sigma}\right) - \Phi\left(-\frac{a}{\sigma}\right) =
                                \mathrm{erf}\left(\frac{a}{\sigma\sqrt{2}}\right) .
\end{equation}
The correlator responses to two identical signals, $\hat{x}^2$, make the series of squares $h_1^2, h_2^2, \ldots, h_{m-1}^2, h_m^2$. Looking at Fig.~\ref{odd_erfs_levels} one can notice that the probability of occurrence of each squared level (except zero) equals to the areas of two curvilinear trapezoids symmetrical with respect to the $\hat{x}$ axis. Consider the sequence of $\hat{x}$-symmetrical, "matryoshka"-nested areas \eqref{prob_sym}: $\Psi(a_1), \Psi(a_2), \ldots, \Psi(a_{m-1}), \Psi(a_m)$, where only thresholds on the positive side are counted from 1 to $m$. The area of a trapezoid pair is the difference between the adjacent couple of the areas in the sequence. For example, the probability of the correlator output $h_1^2$ is $\Psi(a_2) - \Psi(a_1)$. The variance $\hat{\sigma}^2$ of quantized signal $\hat{x}$ is the expectation of correlator response. It is the sum of products of its magnitudes and their probabilities for all the levels of $\hat{x}$ from $h_1$ to $h_m$,
\begin{align}
  \label{variance_terms_odd}
  \hat{\sigma}^2 = 0^2\cdot\Psi(a_1) &+ h_1^2\left[\Psi(a_2)-\Psi(a_1)\right] + \ldots \nonumber \\
           &+ h_{m-1}^2\left[\Psi(a_m)-\Psi(a_{m-1})\right] + h_m^2\left[1-\Psi(a_m)\right],
\end{align}
or, in a more concise fashion,
\begin{equation} 
  \label{sig_as_fun_qsig_odd}
  \hat{\sigma} = \left[ h_m^2 - h_1^2\Psi(a_1) - \sum\limits_{k=2}^m \left[ h_k^2 - h_{k-1}^2 \right] \Psi(a_k) \right]^{1/2}.
\end{equation} 

\begin{figure}[ht] 
\noindent\includegraphics[width=30pc]{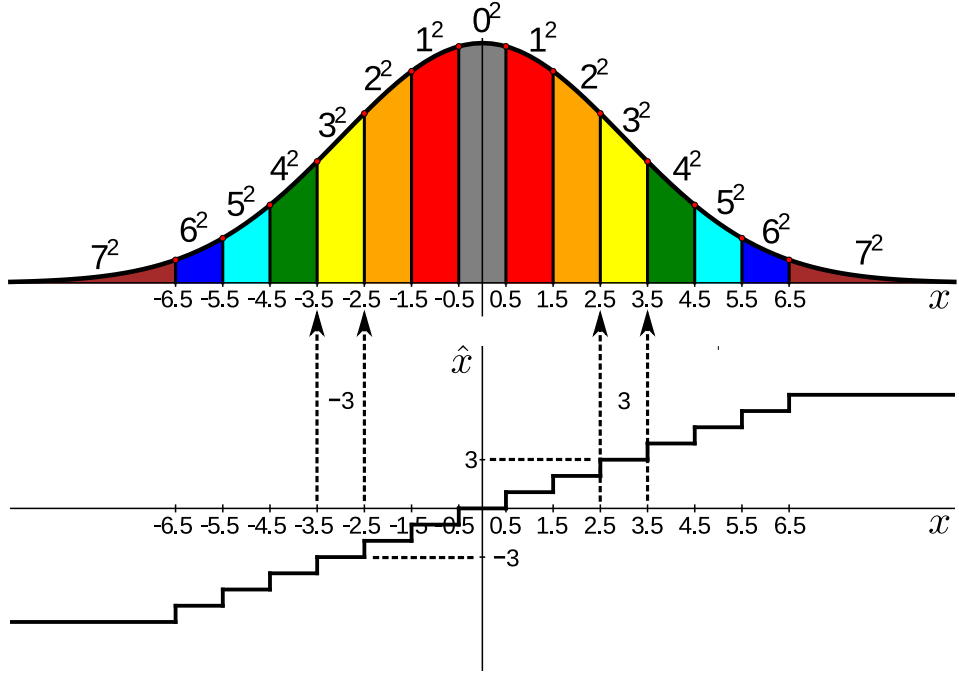}
\caption{\small  Normal probability density law over the intervals of equal correlator output for the 4-bit ADC pattern, used, in particular, in MWA. The equal output areas are filled with the same color.}
\label{erfs_levels}
\end{figure}
Note that for quantizers without the DC response, like those with the patterns shown in panels (a) and (c) of Fig.~\ref{fig_simple_qpatterns}, Eq.~\eqref{sig_as_fun_qsig_odd} is also applicable, if the first switching threshold is set to zero: $a_1=0$.

For a regular quantization pattern like that of MWA depicted in Fig.~\ref{adc_4bit_characteristics} function~\eqref{qsig_general} can be made even simpler. Two identical signals make the correlator output integer squares $0, 1, 4, 9 \ldots, m^2$, and the probability of occurrence of each except zero is sum of the areas of two curvilinear trapezoids symmetrical with respect to the $\hat{x}$ axis, as shown in Fig.~\ref{erfs_levels}. Following the previous example, we write down the variance $\hat{\sigma}^2$ of quantized signal $\hat{x}$ as the sum of products of the correlator responses and their probabilities for all the levels of $\hat{x}$ from 1 up to $m=7$ (see Eq.~\eqref{maxlevel}):
\begin{equation}
  \label{variance_terms_4bit}
  \hat{\sigma}^2 = 0^2\cdot\Psi(0.5) + 1^2\cdot\left[\Psi(1.5)-\Psi(0.5)\right] + 
           2^2\cdot\left[\Psi(2.5)-\Psi(1.5)\right] + ... + 7^2\cdot\left[1-\Psi(6.5)\right].
\end{equation}
Expansion of the terms and their regrouping leads to the compact formula for the quantized STD estimate
\begin{equation} 
  \label{sig_as_fun_qsig_regular}
  \hat{\sigma} = \left[ m^2 - \sum\limits_{k=0}^{m-1} (2k+1)\Psi(k+0.5) \right]^{1/2},
\end{equation} 
where $m$ is the maximal value for the $N$-bit integer, given in Eq.~\eqref{maxlevel}. For the MWA case of the $N=4$-bit integers, $m=7$. 

Eqs.~\eqref{qsig_general}, \eqref{sig_as_fun_qsig_odd}, and \eqref{sig_as_fun_qsig_regular} are  representations of the statistical link~\eqref{qsig_as_fun_sig} between the analog STD and the STD of the same signal after a quantization.

\section{Implementation of the Van Vleck correction and simulation results}

\begin{figure}[ht] 
\noindent\includegraphics[width=25pc]{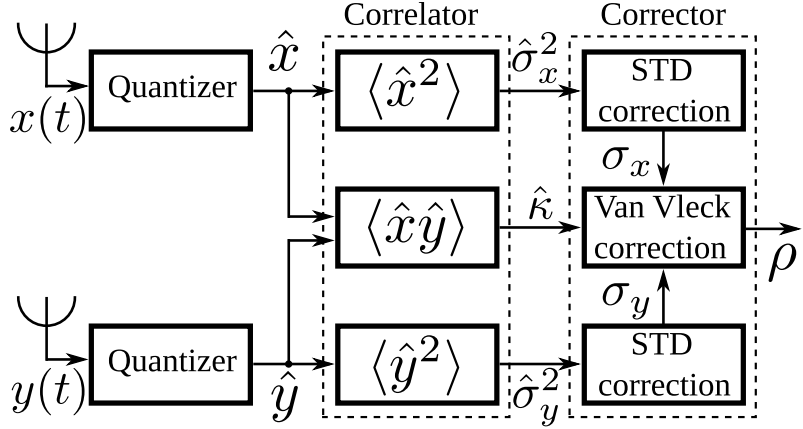}
\caption{\small Implementation of the Van Vleck correction of the covariance of two real signals. For each two quantized antenna signals, $\hat{x}$ and $\hat{y}$, the correlator stage generates three products: the quantized covariance, $\hat{\kappa}$, and the quantized variances, $\hat{\sigma}_x^2$ and $\hat{\sigma}_y^2$. The Van Vleck correction block uses Eq.~\eqref{qkappa_as_integral} to find the analog correlation coefficient $\rho$. The block requires three inputs: quantized covariance $\hat{\kappa}$ and two analog standard deviations, $\sigma_x$ and $\sigma_y$. The analog STDs are estimated with the use of Eq.~\eqref{qsig_general} from their quantized counterparts $\hat{\sigma}_x$ and $\hat{\sigma}_y$.}
\label{std_and_corr_correction}
\end{figure}

Now that we have the methods to estimate both standard deviations of two signals and their correlation, we can outline a practical scheme for the Van Vleck correction, as shown in Fig.~\ref{std_and_corr_correction}. For each antenna pair the analog signals $x(t)$ and $y(t)$ with generally unequal RMS values $\sigma_x$ and $\sigma_y$ are quantized using an arbitrary quantization pattern (see, for example, Fig.~\ref{gen_qpattern} and Eq.~\eqref{qpattern}) into the streams of binary numbers $\hat{x}$ and $\hat{y}$. In real multi-channel systems, $\hat{x}$ and $\hat{y}$ can be a pair of narrow-band channels at the output of a polyphase filter bank (PFB) sampled at the Nyquist rate. For each signal pair the correlator stage generates their quantized covariance, $\hat{\kappa}$, and the quantized  variances, $\hat{\sigma}_x^2$ and $\hat{\sigma}_y^2$. In order to find the corrected correlation coefficient $\rho$ with the use of Eq.~\eqref{qkappa_as_integral} three values must be known: the quantized covariance $\hat{\kappa}$ and two analog standard deviations, $\sigma_x$ and $\sigma_y$. While $\hat{\kappa}$ is the correlator output, the other two need to be estimated from their quantized values using Eq.~\eqref{qsig_general}. 

Modern radio interferometers use the two-component, complex signal representation, called ``analytic signal''. The covariance $\hat{\kappa}$ of analytic signals is also a complex quantity. The components of the complex analytic signal, $I(t)+\mathrm{j}Q(t)$, are called ``in-phase'', $I(t)$, and ``quadrature'', $Q(t)$. The quadrature component, $Q(t)$, is the Hilbert transform \citep[p. 267]{Bracewell1986} of the in-phase component, $I(t)$, i.e. the result of $90^\circ$ phase shift of all the spectral components of $I(t)$ with their amplitudes unchanged. As a consequence, all the stochastic analytic signals belong to the important class of circularly symmetric processes. A complex random variable $z$ is circularly symmetric if it has zero mean and the result of its rotation by an arbitrary angle $\theta$, $w = e^{\uj \theta}z$, has the same distribution as $z$. 

We denote quantized analytic signals from two antennas, $i$-th and $k$-th, as $\hat{z}_{il}$ and  $\hat{z}_{kl}$: 
\begin{eqnarray} 
  \label{quant_signals}
  \hat{z}_{il} &=& \hat{x}_{il} + \uj \, \hat{y}_{il}, \nonumber \\
  \hat{z}_{kl} &=& \hat{x}_{kl} + \uj \, \hat{y}_{kl},
\end{eqnarray}
where the integer index $l$ is for the sample numbers (to be omitted further). The correlator response for each antenna pair is thus
\begin{equation}
  N\hat{\kappa} = \sum\limits_l \hat{z}_{il}\hat{z}_{kl}^*,
\end{equation}
where the asterisk ``$^*$'' means the complex conjugation. Expansion of $\hat{\kappa}$ yields
\begin{equation}
  \label{cplx_kappa}
  \hat{\kappa} = [\langle \hat{x}_i\hat{x}_k \rangle + \langle \hat{y}_i\hat{y}_k \rangle] 
  		   + \uj [-\langle \hat{x}_i\hat{y}_k \rangle + \langle \hat{y}_i\hat{x}_k \rangle].
\end{equation}
Due to the circular symmetry of $\hat{z}_i$ and $\hat{z}_k$, the covariances of the terms of their complex components are equal:
\begin{eqnarray} 
  \label{equal_comps_of_cov}
  \hat{\kappa}_\mathrm{re} =& \langle \hat{x}_i\hat{x}_k \rangle \equiv \langle \hat{y}_i\hat{y}_k \rangle, \nonumber \\
  \hat{\kappa}_\mathrm{im} =& -\langle \hat{x}_i\hat{y}_k \rangle \equiv \langle \hat{y}_i\hat{x}_k \rangle,
\end{eqnarray}
and the complex covariance $\hat{\kappa}$ can be rendered as
\begin{equation}
  \label{cplx_kre_kim}
  \hat{\kappa} = \Real[\hat{\kappa}] + \uj \, \Imag[\hat{\kappa}] 
               = 2(\hat{\kappa}_\mathrm{re} + \uj \hat{\kappa}_\mathrm{im}).
\end{equation}
The halved components of the quantized complex covariance $\hat{\kappa}$,
\begin{equation}
  \label{kre_kim}
  \hat{\kappa}_\mathrm{re} = \tfrac{1}{2}\Real[\hat{\kappa}]  \;\; \mathrm{and} \;\;  
  \hat{\kappa}_\mathrm{im} = \tfrac{1}{2}\Imag[\hat{\kappa}],
\end{equation}
are used in the Van Vleck correction procedure for obtaining the components of the true analog correlation,
\begin{equation}
  \label{cplx_correlation}
  \rho = \rho_\mathrm{re} + \uj \, \rho_\mathrm{im}.
\end{equation}
Note that the variance $\sigma_z^2$ of a circularly symmetric complex random variable $z = x + \uj y$ is two times larger than the variances $\sigma_x^2$ and $\sigma_y^2$ of its individual components, which are equal in the circularly symmetric case,
\begin{equation} \label{cplxvar}
  \sigma_{z}^2 = \langle x^2 + y^2 \rangle = \langle x^2 \rangle + \langle y^2 \rangle = \sigma_{x}^2 + \sigma_{y}^2 =
      2\sigma^2,
\end{equation}
so the standard deviation of the circularly symmetric complex signal and those of its real components are related as follows:
\begin{equation} \label{cplx_vs_real_std}
  \sigma_{z} = \sqrt{2}\sigma_x = \sqrt{2}\sigma_y = \sqrt{2}\sigma.
\end{equation}
In order to correct the component covariances $\hat{\kappa}_\mathrm{re}$ and $\hat{\kappa}_\mathrm{im}$ we need to obtain the component standard deviation from the complex variance, $\hat{\sigma}_z^2$,  found in the observation data set:
\begin{equation}
  \label{comp_std}
  \hat{\sigma} = \sqrt{\hat{\sigma}_z^2/2}.
\end{equation}  
One can notice that the correction function (Eq.~\ref{rho_as_fun_qkappa}) relates the quantized covariance of \emph{real} signals to the analog correlation coefficient $\rho$, while the components of the complex covariance \eqref{cplx_kappa} are doubled \emph{real} covariances, as shown in Eqs.~\eqref{equal_comps_of_cov} to~\eqref{kre_kim}, Hence, the correction function relates the halved value of a component of complex quantized covariance $\hat{\kappa}$ to the corresponding component of the complex correlation $\rho$. Therefore, the procedure of \emph{complex} Van Vleck correction is reduced to applying the correction function~\eqref{rho_as_fun_qkappa} to two \emph{halved} components of the complex covariance $\hat{\kappa} = 2(\hat{\kappa}_\mathrm{re} + \mathrm{j} \hat{\kappa}_\mathrm{im})$ individually:
\begin{eqnarray}
  \label{corr_fun_cplx}
  \rho_\mathrm{re} &= g^{-1}(\hat{\kappa}_\mathrm{re},\sigma_i,\sigma_k), \nonumber \\
  \rho_\mathrm{im} &= g^{-1}(\hat{\kappa}_\mathrm{im},\sigma_i,\sigma_k).
\end{eqnarray}
This reasoning does not use the property that the correlated signals are analytic, i.e. have no negative frequency in their spectra, or the imaginary components are the Hilbert transforms of their real ones. This correction method is applicable to the more general class of complex-valued signals: the only restriction is that the signals must be circularly symmetric complex stochastic processes sampled from the Gaussian distributions. 

\begin{figure}[ht] 
\noindent\includegraphics[width=32pc]{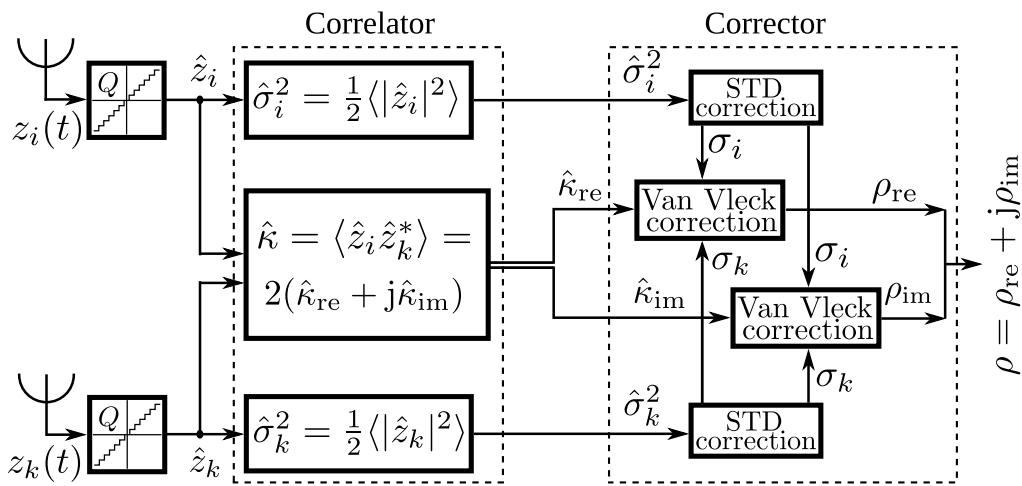}
\caption{\small Implementation of the Van Vleck correction of the covariance of two complex signals. The signals from antenna $i$, $z_i(t)$, and antenna $k$, $z_k(t)$, are quantized (in the Q-blocks) into $\hat{z}_i$ and $\hat{z}_k$. The correlator stage generates their quantized cross-covariance, $\hat{\kappa}$  and their quantized variances, $\hat{\sigma}_i^2$ and $\hat{\sigma}_k^2$. As shown in Eqs.~\eqref{cplx_kappa}~to~\eqref{kre_kim}, the complex quantity $\hat{\kappa}$ has two components, $\hat{\kappa}_\mathrm{re}$ and $\hat{\kappa}_\mathrm{im}$. In the corrector block the analog STDs are estimated with the use of Eq.~\eqref{qsig_general} from their quantized counterparts $\hat{\sigma}_i$ and $\hat{\sigma}_k$ and used for the Van Vleck correction. Each of the $\hat{\kappa}$ components is corrected individually (as a real value) using Eq.~\eqref{qkappa_as_integral} to produce the components of complex analog correlation coefficient, $\rho_\mathrm{re}$ and $\rho_\mathrm{im}$.}
\label{std_and_corr_correction_cplx}
\end{figure}

A schematics of the complex Van Vleck correction is presented in Fig.~\ref{std_and_corr_correction_cplx}. The correction functions $\rho = g^{-1}(\hat{\kappa},\sigma_x,\sigma_y)$ (Eq. \ref{rho_as_fun_qkappa}) and $\sigma = f^{-1}(\hat{\sigma})$ (Eq. \ref{sig_as_fun_qsig}) have been implemented as tables with efficient lookup algorithms. Because this work is focused on the 4-bit MWA correlator, only regular quantization patterns were considered, such as those shown in Fig.~\ref{adc_4bit_characteristics} and in plates (b) and (d) of Fig.~\ref{fig_simple_qpatterns}, so we used Eqs.~\eqref{qkappa_as_integral_mwa} and~\eqref{sig_as_fun_qsig_regular}.

\begin{figure}[ht] 
\noindent\includegraphics[width=25pc]{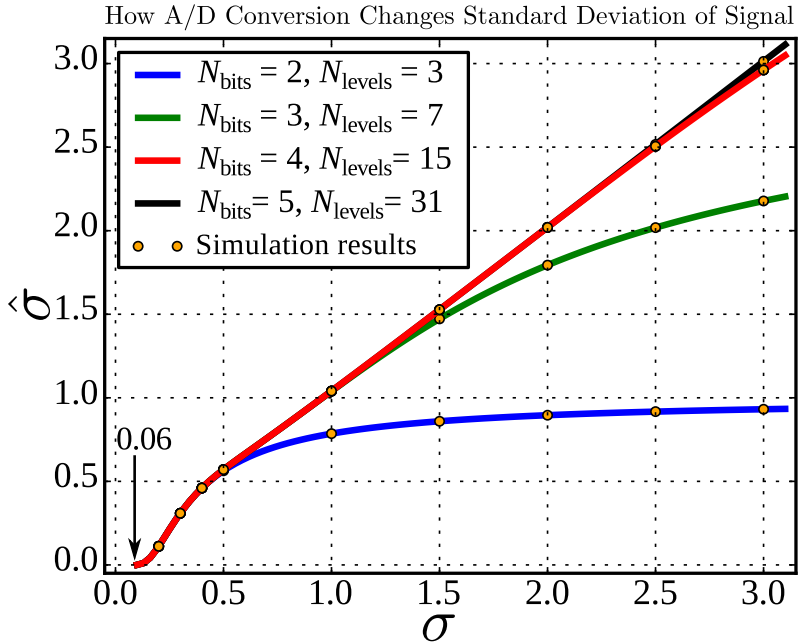}
\caption{\small  Theoretically predicted standard deviations of the quantized signals on ADC output as functions of the standard deviations of the ADC analog inputs. }
\label{qstd_vs_std}
\end{figure}

In order to determine the $f^{-1}(\hat{\sigma})$ table parameters, its extent and fineness, behavior of the function was tested. For ADC precisions of 2, 3, 4, and 5 bits the function $\hat{\sigma} = f(\sigma)$ is plotted in Fig.~\ref{qstd_vs_std}. The curves show the predictions of how the standard deviation of an ADC digital output depends on the standard deviation of its analog input. Also, we made simulations of the Gaussian signal quantization. We calculated the standard deviations of several samples of $10^7$ normally distributed, double precision random numbers. These samples then were quantized with integer precision of 2, 3, 4, and 5 bits, and the standard deviations of the quantized samples were plotted over the theoretical curves, showing excellent agreement. One can see that for the 4-bit quantization in the range of interest, when $\sigma$ is between 0.5 and 2.0, the error between $\sigma$ and $\hat{\sigma}$ is substantial, especially for the smaller $\sigma$'s. Even for $\sigma=1.0$, the 4-bit quantization infers $\hat{\sigma}=1.041$, i.e. the error is about 4\%. The lower limit for input analog RMS is $0.06$: the RMS of an ADC response to analog signals below $\sigma=0.06$ is zero. We created a table of $\hat{\sigma}$ values with the key column (or ruler) of $\sigma$ from 0.06 to 3.6 with the increment of 0.02. The table contains values for 2, 3, 4, and 5-bit ADCs, computed from Eq.~\eqref{sig_as_fun_qsig_regular}.

\begin{figure}[ht] 
\noindent\includegraphics[width=20pc]{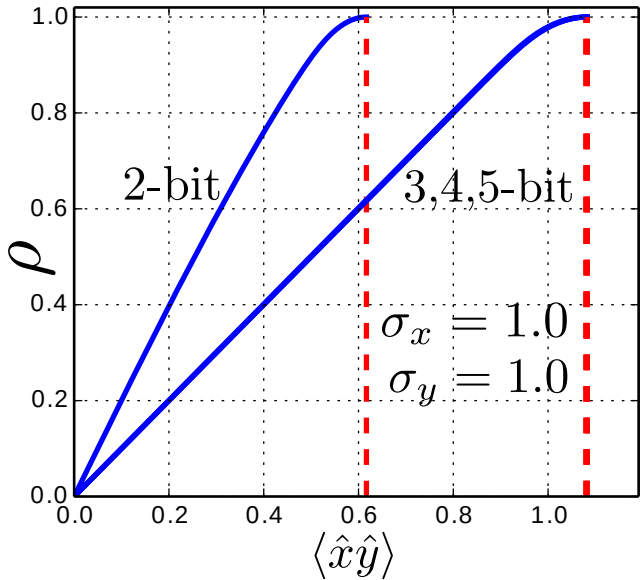}
\caption{\small  The Van Vleck correction curves $\rho = g^{-1}(\hat{\kappa},\sigma_x,\sigma_y)$, where $\hat{\kappa}$ is the covariance $\langle\hat{x}\hat{y}\rangle$, for 2, 3, 4, and 5-bit ADCs and equal STDs of the signals, $\sigma_x=1,\sigma_y=1$. The difference between the 3, 4, and 5-bit curves is invisible. All the curves have convex ends. }
\label{vanvleck_2345b_s11}
\end{figure}

\begin{figure}[ht] 
\noindent\includegraphics[width=20pc]{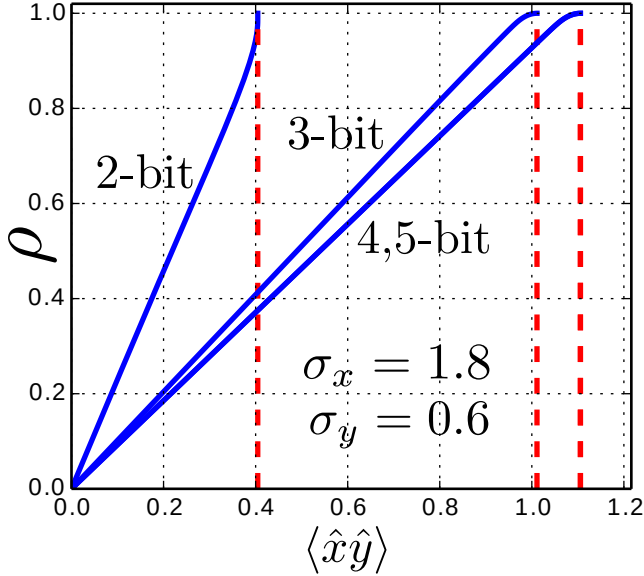}
\caption{\small  The Van Vleck correction curves $\rho = g^{-1}(\hat{\kappa},\sigma_x,\sigma_y)$, where $\hat{\kappa}$ is the covariance $\langle\hat{x}\hat{y}\rangle$, for 2, 3, 4, and 5-bit ADCs and different STDs of the signals, $\sigma_x=1.8,\sigma_y=0.6$. The 4- and 5-bit curves are very close and indistinguishable. While the 3, 4, and 5-bit curves have convex ends, the tip of the 2-bit curve is concave.}
\label{vanvleck_2345b_s1.8_0.6}
\end{figure}

\begin{figure}[ht] 
\noindent\includegraphics[width=20pc]{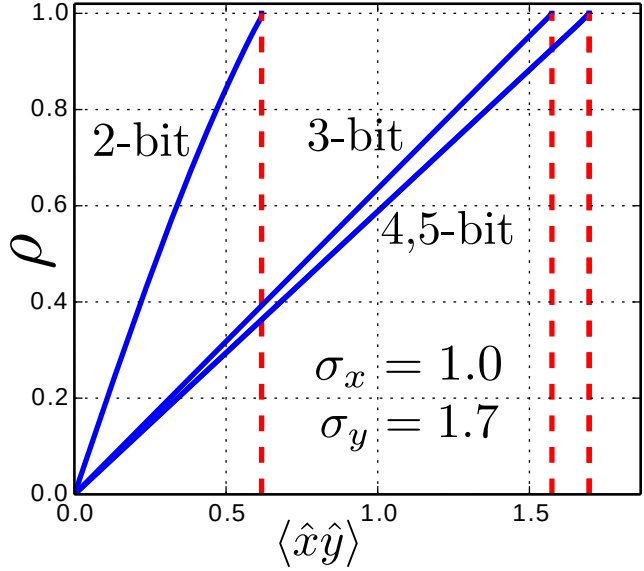}
\caption{\small  The Van Vleck correction curves $\rho = g^{-1}(\hat{\kappa},\sigma_x,\sigma_y)$, where $\hat{\kappa}$ is the covariance $\langle\hat{x}\hat{y}\rangle$, for 2, 3, 4, and 5-bit ADCs and different STDs of the signals, $\sigma_x=1.0,\sigma_y=1.7$. The 4- and 5-bit curves are very close and indistinguishable. }
\label{vanvleck_2345b_s1_1.7}
\end{figure}

Several sample graphs of the Van Vleck correction function $\rho = g^{-1}(\hat{\kappa},\sigma_x,\sigma_y)$ (Eq. \ref{rho_as_fun_qkappa}) are shown in Figs.~\ref{vanvleck_2345b_s11}, \ref{vanvleck_2345b_s1.8_0.6}, and \ref{vanvleck_2345b_s1_1.7}. The function was implemented for the 4-bit MWA quantizer as a three-dimensional table. The first two dimensions, $\sigma_1$ and $\sigma_2$, have finer rulers than the key column of the $\sigma = f^{-1}(\hat{\sigma})$ table. Both dimensions are from 0.06 to 3.2 with the step 0.01. The ruler of third dimension of the table, $\rho$, has 101 number rising from 0 to $0.9999999=1-10^{-7}$. The values of $\rho$ obey the logarithmic scale, because functional dependence \eqref{rho_as_fun_qkappa} with respect to $\hat{\kappa}$ is almost linear everywhere but the close vicinity of $|\rho|\approx 1$. Only the positive values of $\rho$ are used because its dependence on $\hat{\kappa}$ is odd.

Both $\sigma$ and $\rho$ tables use the bisection search for fast (in $\sim\log_2n$ tries) finding the place in the table and subsequent linear interpolation. However, in the $\rho$ table each bisection step uses the bilinear interpolation over the $(\sigma_x,\sigma_y)$ grid. The methods are described in Numerical Recipes in Fortran by \citet[sec. 3.4, pp. 110-111; sec. 3.6, pp 116-117]{Numrecp1992} 

\begin{figure}[ht] 
\noindent\includegraphics[width=20pc]{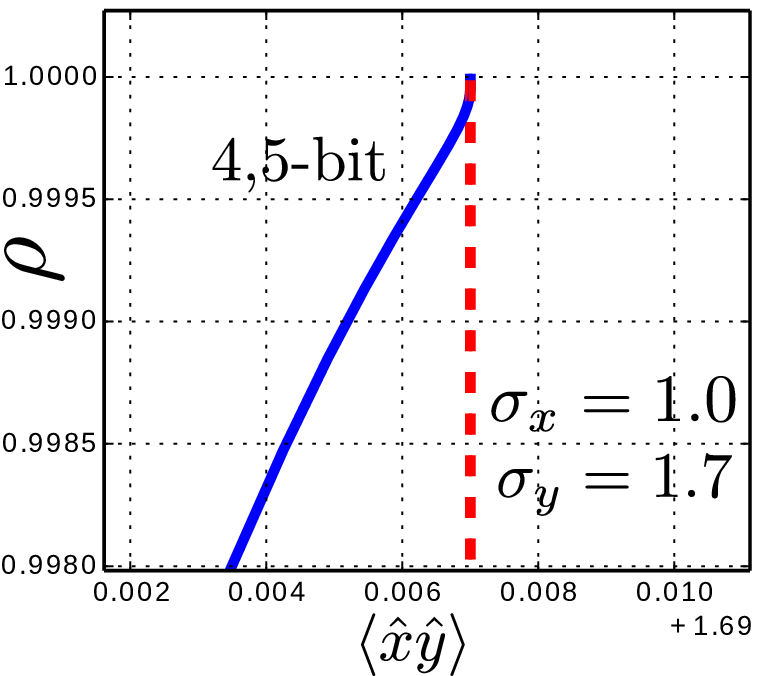}
\caption{\small  A zoomed-in view of the curves in previous Fig.~\ref{vanvleck_2345b_s1_1.7}. The upper tips of the Van Vleck correction curves $\rho = g^{-1}(\hat{\kappa},\sigma_x,\sigma_y)$, where $\hat{\kappa}$ is the covariance $\langle\hat{x}\hat{y}\rangle$, for 4 and 5-bit ADCs and different STDs of the signals, $\sigma_x=1.0,\sigma_y=1.7$. The curves are very close and indistinguishable. Near $\rho\approx1$ the curve changes from convex to concave. }
\label{vanvleck_2345b_s1_1.7_tips}
\end{figure}

\begin{figure}[ht] 
\noindent\includegraphics[width=25pc]{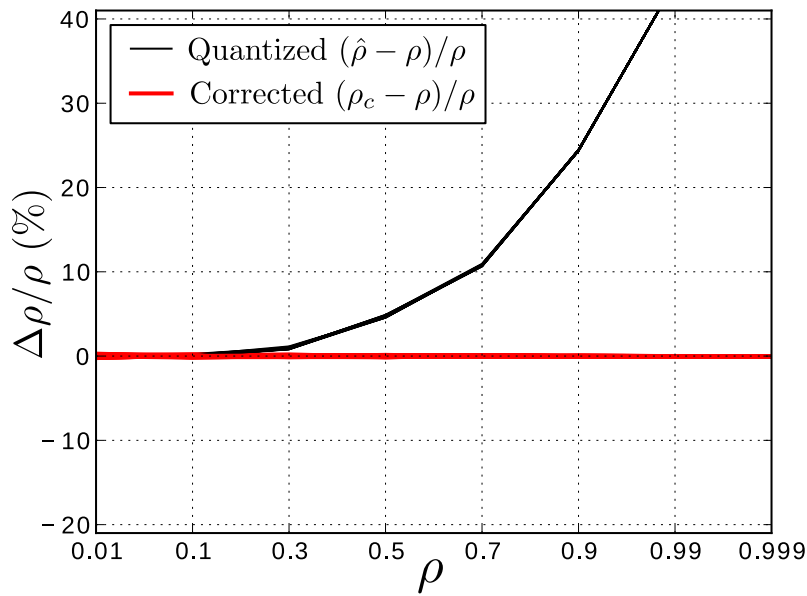}
\caption{\small  Relative errors of the 2-level quantized correlation $\hat{\rho}$ before and after the Van Vleck correction as dependent on the correlation level. }
\label{fig_relerr_2lev}
\end{figure}

\begin{figure}[ht] 
\noindent\includegraphics[width=25pc]{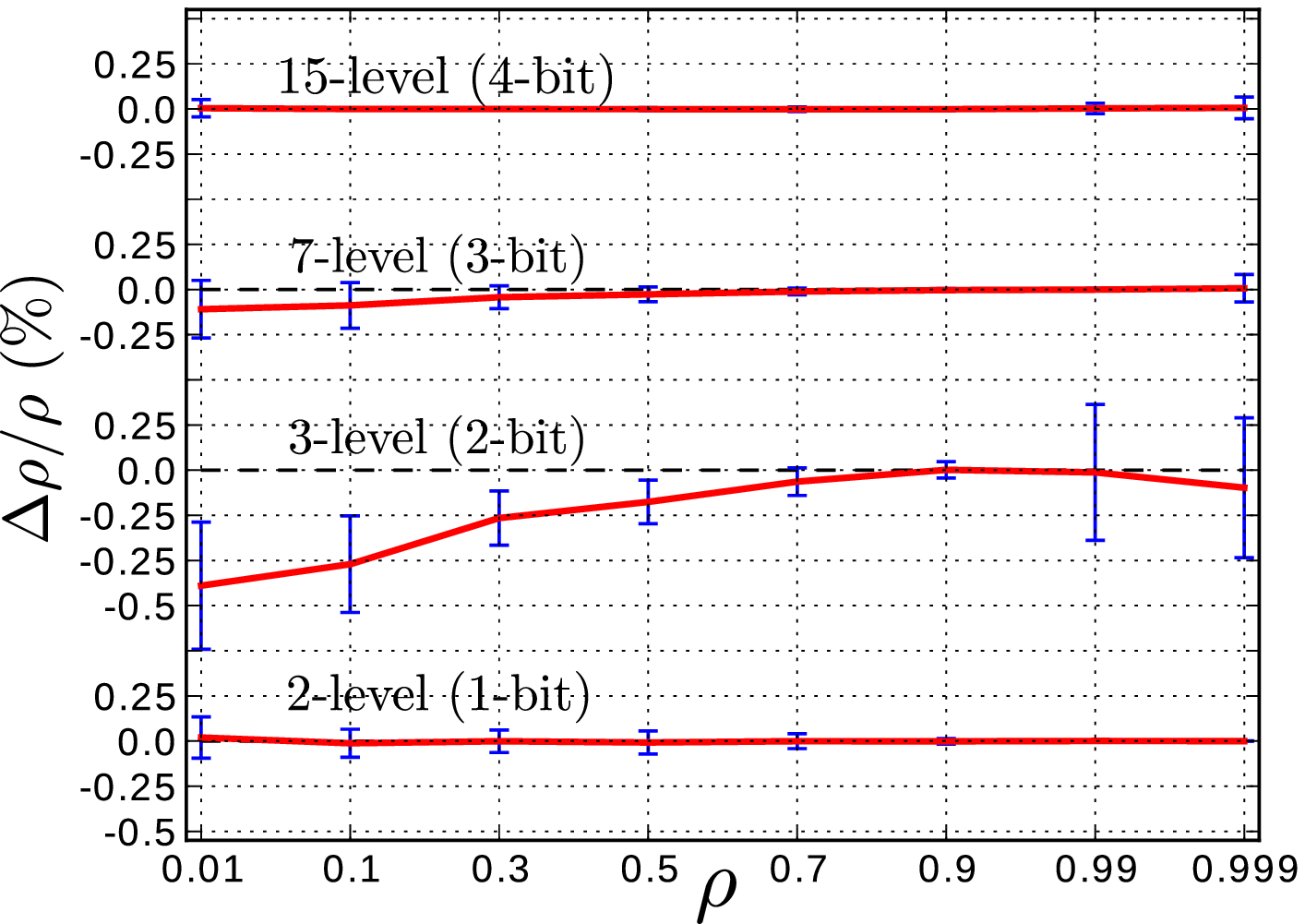}
\caption{\small  Means and uncertainties of the relative errors of the Van Vleck corrected correlations $\rho_c$ for 2, 3, 7, and 15 quantization levels as dependent on the correlation level. }
\label{fig_means_errorbars_all_lev}
\end{figure}

A series of simulations for the purpose of testing the Van Vleck correction method have been conducted, demonstrating good results. Utilizing an Nvidia multicore graphical processor (GPU) with the CUDA software increased the correlation computation speed by a factor of $\sim$100-200 with respect to that of an Intel i7 PC. This allowed us to simulate analog signals consisting of $10^{10}$ double precision floating point numbers. To imitate the analog signals we used the standard  pseudo-random number generator {\tt curand\_normal\_double()} from the CURAND package. Each signal comprised $10^{10}$ samples drawn from a normal distribution with zero mean and specified standard deviation. The analog signals $x(t)$ and $y(t)$ were composed of two random parts, common signal $s(t)$ and uncorrelated noise, $n_1(t)$ and  $n_2(t)$:
\begin{eqnarray}
\label{xysignals}
  x(t) &=& \sigma_x [a s(t) + b n_1(t)], \nonumber \\
  y(t) &=& \sigma_y [a s(t) + b n_2(t)],
\end{eqnarray}
where $a$ and $b$ were the signal/noise mixing coefficients. The signal-to-noise ratio (SNR) is $q = \mathfrak{R}_\mathrm{SN} = a/b$. Although the method can handle signals with different SNRs, for simplicity in this test we correlated the signals with equal SNRs. In order to find $a$ and $b$ mixing the signal $s$ and noise $n$ in a required proportion $q$, we assumed $s$, $n$, and their mixture $x=as+bn$ having variances equal to unity: $\sigma_{as+bn}^2 = \sigma_s^2 = \sigma_n^2 = 1$. Then $\sigma_{as+bn}^2 = \langle (as+bn)^2 \rangle = 1$. In the binomial expansion the uncorrelated products do not survive, so $\langle (as+(a/q)n)^2 \rangle = a^2 \left( \langle s^2 \rangle + (1/q^2) \langle n^2 \rangle \right) = 1$. Since both variances, $\langle s^2 \rangle$ and $\langle n^2 \rangle$, were set to unity, we get $a^2(1+(1/q^2))=1$. Using $b=a/q$ we can express the signal/noise mixing coefficients as functions of the SNR $q$:
\begin{eqnarray}
\label{ab_from_snr}
  a &=& q/\sqrt{1+q^2}, \nonumber \\
  b &=& 1/\sqrt{1+q^2}.
\end{eqnarray}
In order to test the method on signals with a variety of correlation coefficients, specified ahead of time, we need to know what SNRs of the signals secure the required correlation $\rho$. To find the dependence, we write down the covariance $\sigma_x\sigma_y\rho = \langle xy \rangle$ using Eqs.~\eqref{xysignals} as $\langle xy \rangle = \sigma_x\sigma_y \langle (a s + b n_1) (a s + b n_2) \rangle$. The uncorrelated products vanish, the variance $\langle s^2 \rangle = 1$, so what is left is $\langle xy \rangle = \sigma_x\sigma_y a^2$, whence the correlation is simply $\rho = a^2$. Using Eqs.~\eqref{ab_from_snr}, we obtain the dependence of the correlation coefficient on the signal's SNR (provided the SNRs are equal for both signals):
\begin{equation}
\label{rho_from_snr}
  \rho = q^2/(1+q^2).
\end{equation}
Inversion of the latter yields the dependence of SNR on the correlation:
\begin{equation}
\label{snr_from_rho}
  \mathfrak{R}_\mathrm{SN} = q = \sqrt{\rho/(1 - \rho)}.
\end{equation}
The simulation was done according to the the scheme in Fig.~\ref{std_and_corr_correction}. The analog STDs were estimated using the inverse of Eq.~\eqref{sig_as_fun_qsig_regular}. The correction (i.e. the analog correlation estimation) was made with the use of the inverse of Eq.~\eqref{qkappa_as_integral_mwa}. For comparison, four quantization schemes were considered: 1-bit (2 levels), 2-bit (3 levels), 3-bit (7 levels), and 4-bit (15 levels). All the four had regular quantization patterns. The first was the clipping scheme shown in Fig.~\ref{fig_simple_qpatterns}~(a), the others being subsets of the last, shown in Fig.~\ref{adc_4bit_characteristics}. The means of relative errors and their uncertainties for the four cases are plotted in Fig.~\ref{fig_means_errorbars_all_lev}. Note that the classical clipping scheme in Fig.~\ref{fig_relerr_2lev} is only presented for reference. It does not require the correction method developed here and the correction is made with the simple formula by \citet{Vanvleck1966} $\rho = \sin((\pi/2)\hat{\rho})$.

Sometimes, when the correlations are systematically low (e.g., in VLBI) the Van Vleck correction is not applied, and the quantized correlation is calculated from the quantized covariance as a linear function
\begin{equation}
  \label{rho_k_kappa}
  \hat{\rho} = k \; \hat{\kappa},
\end{equation}
where $k$ is the slope of Van Vleck curve. Figs.~\ref{vanvleck_2345b_s11}, \ref{vanvleck_2345b_s1.8_0.6}, and~\ref{vanvleck_2345b_s1_1.7} show that this is possible because near the coordinate origin the Van Vleck curves are straight lines. This method accurately reproduces the analog correlation for $|\rho| < 0.1$. In our simulations we use Eq.~\eqref{rho_k_kappa} to calculate $\hat{\rho}$.

\begin{figure}[ht] 
\noindent\includegraphics[width=25pc]{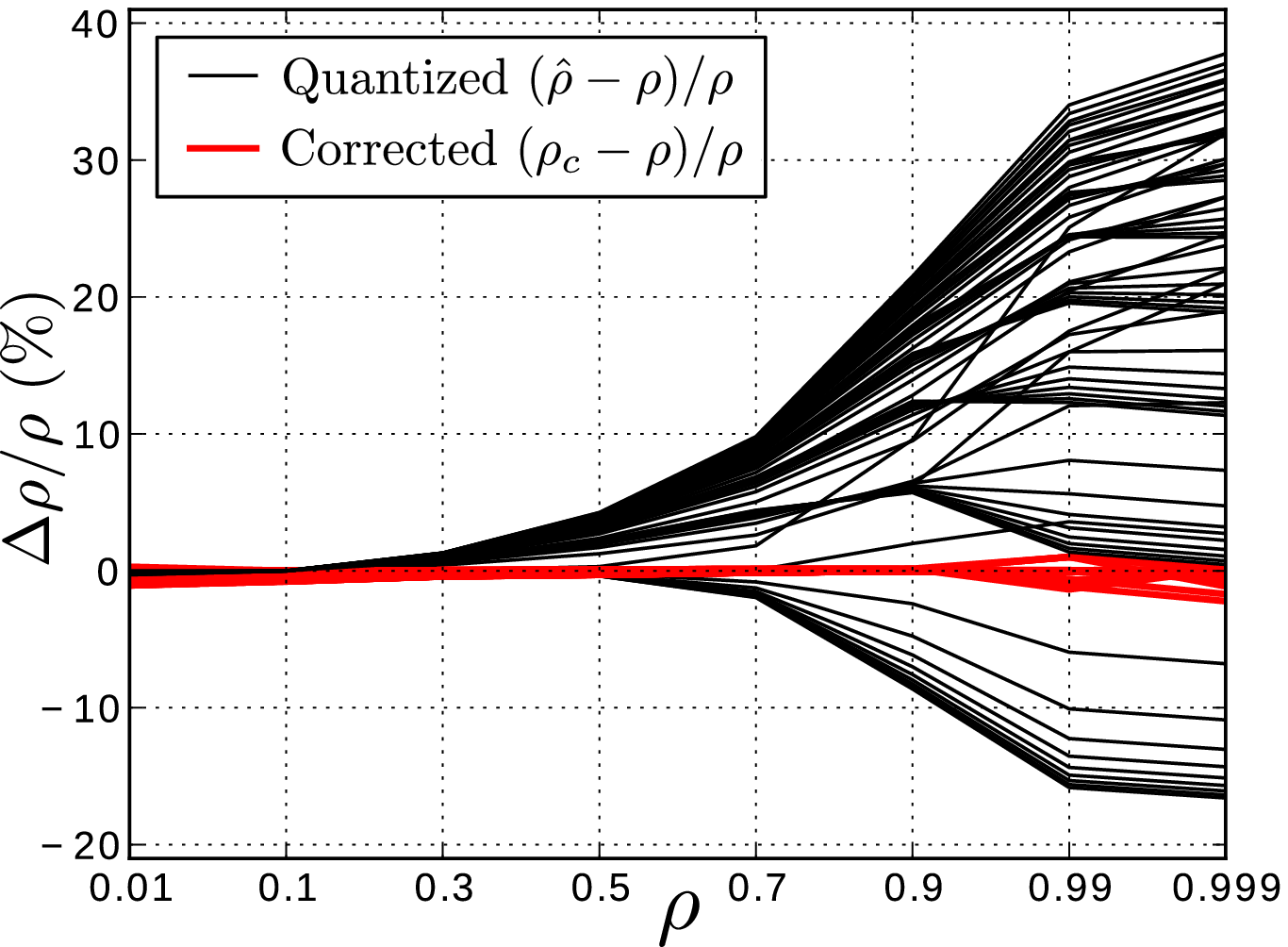}
\caption{\small  Relative errors of the 3-level quantized correlation $\hat{\rho}$ before and after the Van Vleck correction as dependent on the correlation level. }
\label{fig_relerr_3lev}
\end{figure}

\begin{figure}[ht] 
\noindent\includegraphics[width=25pc]{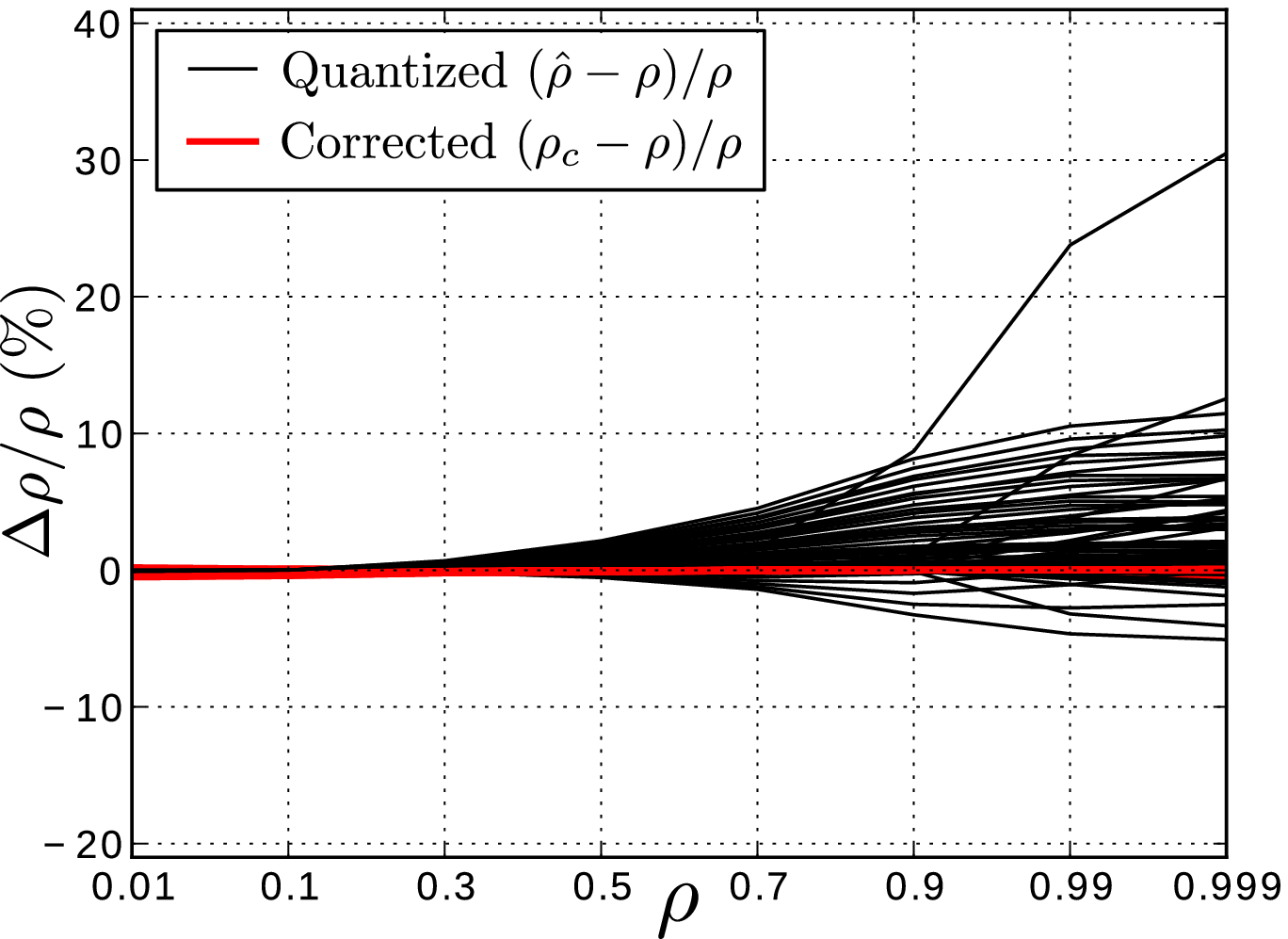}
\caption{\small  Relative errors of the 7-level quantized correlation $\hat{\rho}$ before and after the Van Vleck correction as dependent on the correlation level. }
\label{fig_relerr_7lev}
\end{figure}

\begin{figure}[ht] 
\noindent\includegraphics[width=25pc]{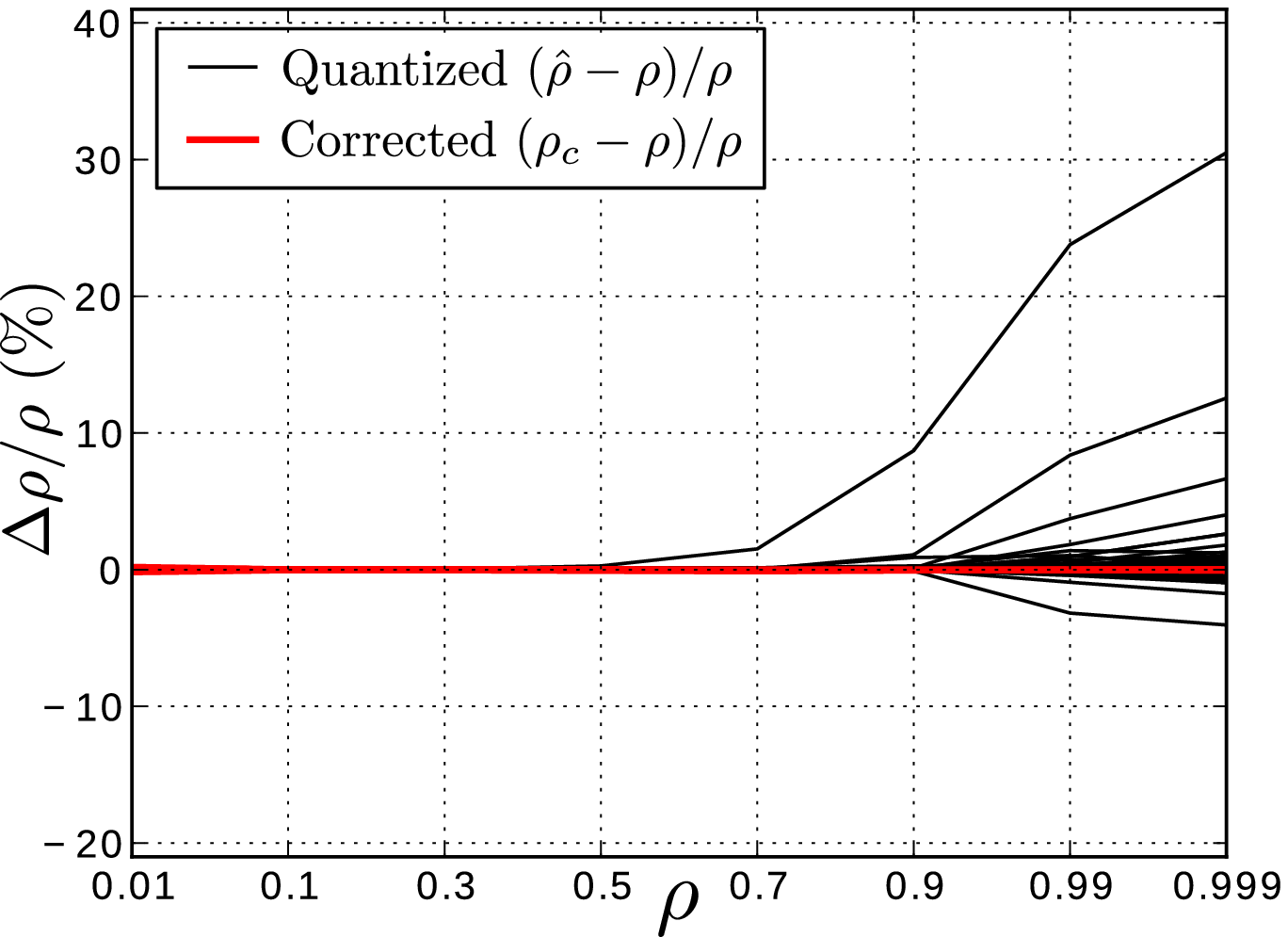}
\caption{\small  Relative errors of the 15-level quantized correlation $\hat{\rho}$ before and after the Van Vleck correction as dependent on the correlation level. }
\label{fig_relerr_15lev}
\end{figure}

We chose a scale of the analog correlation values, $\rho = $ \{0.01, 0.1, 0.3, 0.5, 0.7, 0.9, 0.99, 0.999\}. With the use of Eq.~\eqref{snr_from_rho} we determined the corresponding scale of SNRs: $\mathfrak{R}_\mathrm{SN} = $ \{0.1,  0.33, 0.65,  1.0,  1.53,  3.0, 9.95,  31.6 \}. We correlated the quantized signals with $\sigma_x$ and $\sigma_y$ RMS from the series $\sigma = $ \{0.5,  0.75,  1.,  1.25,  1.5,  1.75,  2.,  2.25,  2.5,  2.75,  3.\}, thus making 66 pairs. For each pair of RMS the dependencies $\hat{\rho}(\rho)$ (quantized correlation), and $\rho_c(\rho)$ (Van Vleck corrected correlation) were obtained. The relative errors for the both, $(\hat{\rho} - \rho)/\rho$ and $(\rho_c - \rho)/\rho$ were plotted versus $\rho$ in Figs.~\ref{fig_relerr_2lev}-\ref{fig_relerr_15lev}.

\section{Discussion}

It is instructive to show that the derived general Van Vleck expression for the covariance of two quantized signals \eqref{qkappa_as_integral} can be used to reproduce the results for a few particular simple Van Vleck correction cases considered, for example, by \citet{TMS2001}.

First is the classical example of the two-level quantizer shown in Fig.~\ref{fig_simple_qpatterns} (a). The quantizer has a single level difference $\Delta h = 2$,  and product of derivatives has only one term, ${\Delta h}^2 = 4$. Since the switching only occurs at zero, the exponent has zero argument. Hence, integral \eqref{qkappa_as_integral} is simplified to 
\begin{equation} 
  \label{qkappa_as_integral_two_level}
   \hat{\kappa} = \frac{1}{2\pi} \int\limits_0^\rho \frac{4 \exp(0)}{\sqrt{1-\zeta^2}}  \mathrm{d}\zeta = 
                  \frac{2}{\pi}  \int\limits_0^\rho \frac{\mathrm{d}\zeta}{\sqrt{1-\zeta^2}}.
\end{equation}
This is a table integral; it immediately gives the familiar Van Vleck relationship
\begin{equation} 
  \label{vanvlec}
  \hat{\kappa} = \frac{2}{\pi}\arcsin\rho.
\end{equation}
Note that the formula calculates the quantized covariance $\hat{\kappa} = \langle \hat{x}\hat{y} \rangle$, which is the correlator output. However, for the two-level quantizer (and only in this case) the quantized covariance is equal to the quantized correlation, $\hat{\rho} = \hat{\kappa}$. In order to get the unknown corrected correlation, $\rho$, it should be inverted:
\begin{equation} 
  \label{vanvlec_rho}
  \rho = \sin\left(\frac{\pi}{2}\hat{\kappa}\right).
\end{equation}

As a second example we shall derive the Van Vleck correction formula for the three-level quantizer, shown in Fig.~\ref{fig_simple_qpatterns}~(b). We follow the notation given in \citet[section 8.3, pp. 271-272]{TMS2001}. Here $n=3$, the levels are $h_1 = -1$, $h_2 = 0$, $h_3 = 1$, the level distances are $\Delta h_1 = \Delta h_2 = 2$, and the switching thresholds are $a_1 = -v_0$ and $a_2 = v_0$. We assume equal variances for both correlated signals $x$ and $y$: $\sigma_x^2 = \sigma_y^2 = \sigma^2$. Substitution of all these into \eqref{qkappa_as_integral} immediately gives us Eq.~(8.55) from \citep{TMS2001}:
\begin{equation} 
  \label{qkappa_as_integral_3lev}
   \hat{\kappa} = \frac{1}{\pi} \int\limits_0^\rho
      \frac{1}{\sqrt{1-\zeta^2}}  \left[ \exp\left(\frac{-v_0^2}{\sigma^2(1-\zeta)}\right)
                                       + \exp\left(\frac{-v_0^2}{\sigma^2(1+\zeta)}\right) \right] \mathrm{d}\zeta.
\end{equation}
The standard deviation $\sigma$ occurring in \eqref{qkappa_as_integral_3lev} is the RMS of analog signals $x$ and $y$. It can be evaluated with the use of Eq.~\eqref{qsig_general}. Substituting $h_1^2=h_3^2=1$ and $h_2^2=0$, we get
\begin{equation} 
  \label{sig_as_fun_qsig_3lev}
  \hat{\sigma}^2 = \Phi\left(-\frac{v_0}{\sigma}\right) + 1 - \Phi\left(\frac{v_0}{\sigma}\right) = 1 - \Psi(v_0),
\end{equation}
where $\Psi$ is given by Eq.~\eqref{prob_sym}, and $\Phi$ is CDF of the standard normal distribution given by Eq.~\eqref{normal_cdf}. The non-linear Eq.~\eqref{sig_as_fun_qsig_3lev} should be solved for the analog STD $\sigma$ using the known $\hat{\sigma}$. In turn, $\sigma$ is used to solve Eq.~\eqref{qkappa_as_integral_3lev} for the analog correlation coefficient $\rho$, as described in the scheme in Fig.~\ref{std_and_corr_correction}. 

The graph of function~\eqref{sig_as_fun_qsig_3lev}, $\hat{\sigma} = f(\sigma)$, for $v_0=1$ is plotted in Fig.~\ref{qstd_vs_std}, the blue curve marked $N_{\mathrm{bits}}=2, N_{\mathrm{levels}}=3$. One can notice that the quantized STD $\hat{\sigma}$ does not reach unity even for the analog signals with RMS as large as $\sigma=3$.

A third example is the four-level quantizer, whose characteristic is shown in Fig.~\ref{fig_simple_qpatterns}~(b). It is considered in the book by \citet[section 8.2, pp. 264-267]{TMS2001}. The number of levels is $n=4$, the levels are $h_1 = -n$, $h_2 = -1$, $h_3 = 1$, and $h_4=n$. The level distances are $\Delta h_1 = \Delta h_3 = n - 1$, $\Delta h_2 = 1$. The switching thresholds are $a_1 = -v_0$, $a_2 = 0$ and $a_3 = v_0$. The variances for both correlated signals $x$ and $y$ are assumed equal, as before: $\sigma_x^2 = \sigma_y^2 = \sigma^2$. After substitution of these parameters, Eq.~\eqref{qkappa_as_integral} takes exactly the same form as Eq.~(8.42) in \citep{TMS2001}:
\begin{align} 
  \label{qkappa_as_integral_4lev}                                       
   \hat{\kappa} = \frac{1}{\pi} \int\limits_0^\rho
      \frac{1}{\sqrt{1-\zeta^2}}  &\left\{ (n-1)^2 \left[ \exp\left(\frac{-v_0^2}{\sigma^2(1-\zeta)}\right)
                                       + \exp\left(\frac{-v_0^2}{\sigma^2(1+\zeta)}\right) \right] \right. \nonumber \\
        &+ \left. 4(n-1) \exp\left(\frac{-v_0^2}{\sigma^2(1-\zeta^2)}\right) + 2 \right\} \mathrm{d}\zeta.
\end{align}
In order to solve this equation for the unknown correlation $\rho$, three variables must be determined: the quantization level $n$, the covariance $\hat{\kappa} = \langle \hat{x}\hat{y} \rangle$, and the analog standard deviation $\sigma$. While $\hat{\kappa}$ is the immediate correlator output, $\sigma$ needs to be found from the correlator output for the same input channel, or from the quantized STD $\hat{\sigma}$. Since the pattern Fig.~\ref{fig_simple_qpatterns}~(b) is odd, we can use Eq.~\eqref{sig_as_fun_qsig_odd}. It requires only positive levels and thresholds, so we have to rename them. The quantizer has no DC level, so the thresholds are $a_1=0$ and $a_2=v_0$.  The positive levels are $h_1 = 1$ and $h_2 = n$, so $m=2$. Using Eq.~\eqref{sig_as_fun_qsig_odd} we first get
\begin{equation} 
  \label{sig_as_fun_qsig_4lev_subs}
  \hat{\sigma}^2 =  n^2 - 1^2\cdot\Psi(v_0) - \left[ n^2 - 1 \right] \Psi(v_0), 
\end{equation} 
and eventually we arrive at the formula resemblant of the denominator of Eq.~(8.43) in \citet[p.~267]{TMS2001}
\begin{equation} 
  \label{sig_as_fun_qsig_4lev}
  \hat{\sigma}^2 =  \Psi(v_0) + n^2 \left[1 - \Psi(v_0) \right] .
\end{equation} 
The unknown STD $\sigma$ occurs in this equation implicitly, and it should be solved numerically.
Thus obtained $\sigma$ is then used to solve Eq.~\eqref{qkappa_as_integral_4lev} for the analog correlation coefficient $\rho$. The scheme in Fig.~\ref{std_and_corr_correction} illustrates this process. 

The Van Vleck correction does not increase the signal-to-noise ratio of the quantized correlation $\mathcal{R}_\mathrm{SN}$. Therefore, it does not improve the quantization efficiency $\eta$, measured as the ratio between $\mathcal{R}_\mathrm{SN}$ and the signal-to-noise ratio of the correlation computed with the unquantized sampling at the Nyquist rate, $\mathcal{R}_\mathrm{SN\infty}$:
\begin{equation} 
  \label{efficiency}
  \eta = \mathcal{R}_\mathrm{SN} / \mathcal{R}_\mathrm{SN\infty}.
\end{equation}
\citet{Thomson2007} provide convenient formulas for the quantization efficiency. See also \citet[sections 8.2 and 8.3, pp. 256-276]{TMS2001}.

\begin{figure}[ht] 
\noindent\includegraphics[width=24pc]{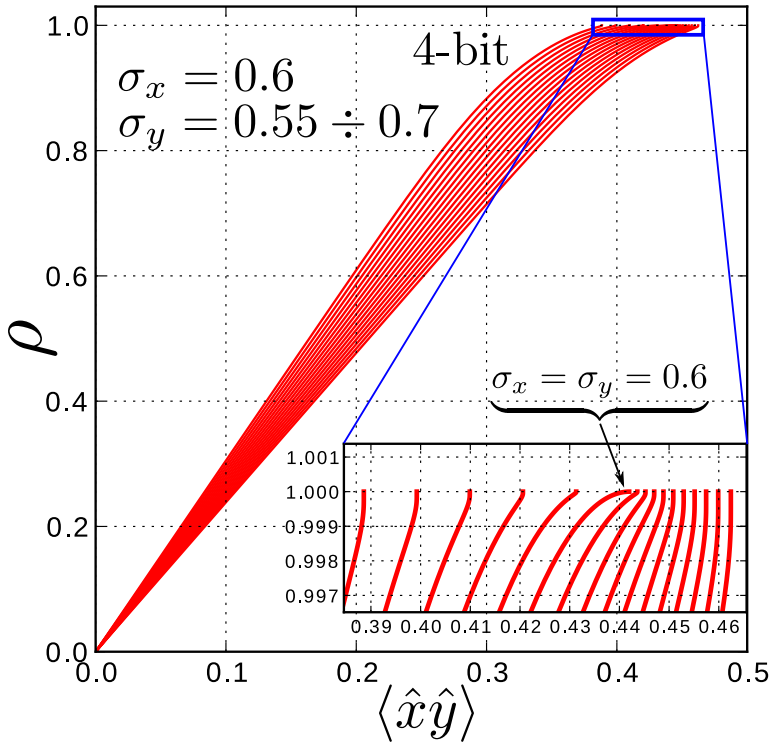}
\caption{\small  The Van Vleck curves $\rho = g^{-1}(\hat{\kappa},\sigma_x,\sigma_y)$ and their upper tips zoomed in (see Eqs. \eqref{rho_as_fun_qkappa}, \eqref{qkappa_as_integral_mwa}, and \eqref{qkappa_as_integral}), for the 4-bit quantizer. The family of 16 curves is plotted for $\sigma_x=0.6$ and $\sigma_y$ varying from 0.55 to 0.7 with a step of 0.01. The zoomed in view corresponds the large analog correlations $|\rho|\approx 1$.  All the curves have concave ends with the exception of one for $\sigma_x=\sigma_y=0.6$. }
\label{vanvleck_s1_0.6_s2_0.55_to_0.7}
\end{figure}

\begin{figure}[ht] 
\noindent\includegraphics[width=20pc]{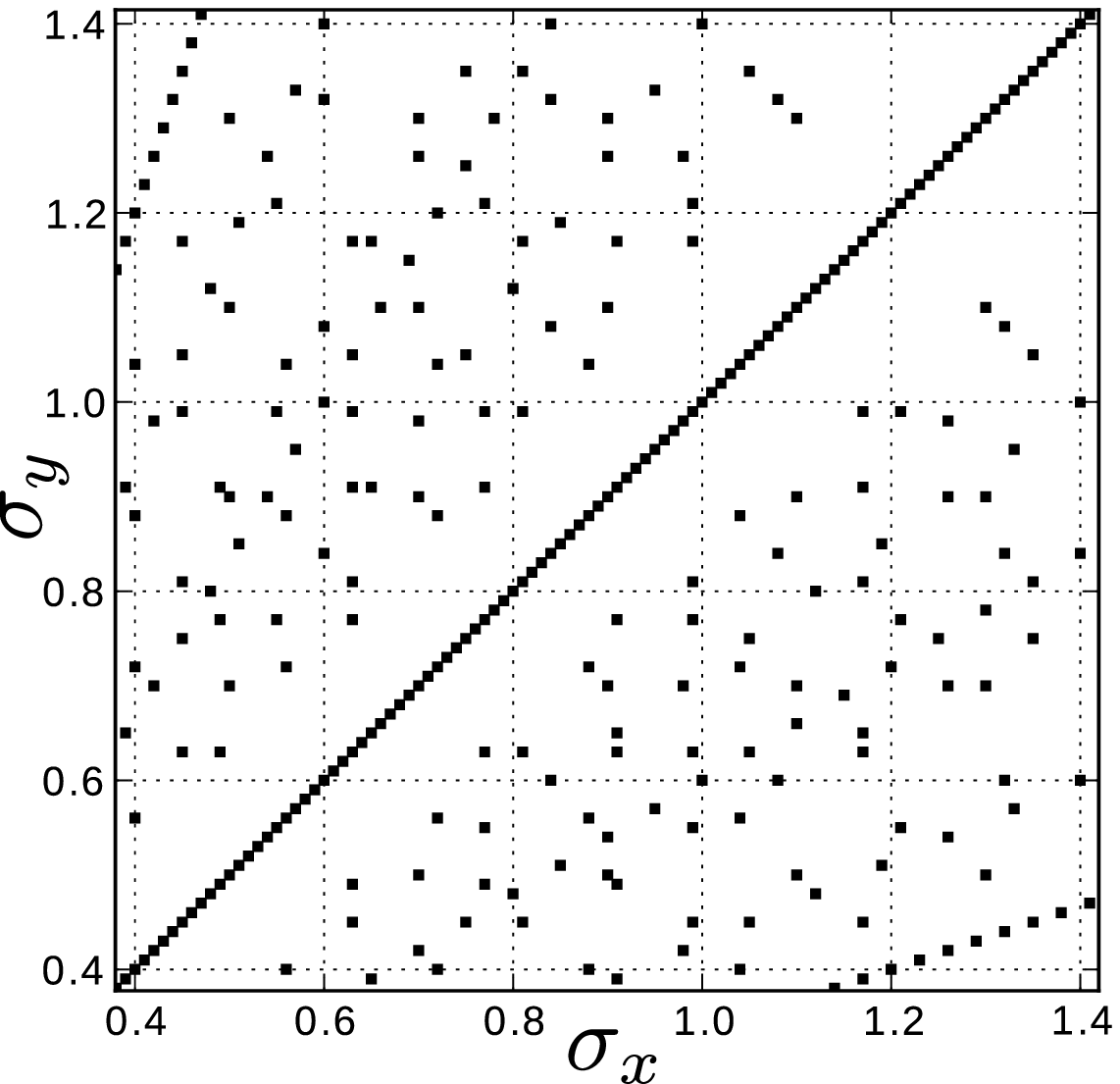}
\caption{\small  Black dots indicate the $(\sigma_x,\sigma_y)$ combinations that result in a convex end of the Van Vleck curve $\rho = g^{-1}(\hat{\kappa},\sigma_x,\sigma_y)$ (see Eqs. \eqref{rho_as_fun_qkappa}, \eqref{qkappa_as_integral_mwa}, and \eqref{qkappa_as_integral}, the 4-bit quantizer) for large analog correlations $|\rho|\approx 1$. The diagonal is filled with the dots because the curves for $\sigma_x = \sigma_y$ always have the convex upper tips. However, the majority of the curves have concave ends.}
\label{where_curve_convex}
\end{figure}

The behavior of Van Vleck correction functions $\rho = g^{-1}(\hat{\kappa},\sigma_x,\sigma_y)$ (Eq. \ref{rho_as_fun_qkappa}) has a curious property: a varying sign of the second derivative with respect to $\hat{\kappa}$ near $|\rho|\approx1$ for different combinations of $\sigma_x$ and $\sigma_y$. In its graphs it looks like convexity (${g^{-1}}^{''}_{\hat{\kappa}}<0$) or concavity (${g^{-1}}^{''}_{\hat{\kappa}}>0$) of the curve's upper tips. For example, in Fig.~\ref{vanvleck_2345b_s11}, where $\sigma_x=\sigma_y=1$, all the curves are convex. In Fig.~\ref{vanvleck_2345b_s1.8_0.6}, where $\sigma_x=1.8$ and $\sigma_y=0.6$, the 2-bit Van Vleck curve is concave, and 3-, 4- and 5-bit curves are convex. In Fig.~\ref{vanvleck_2345b_s1_1.7}, where $\sigma_x=1$ and $\sigma_y=1.7$, it is difficult to make conclusions on the convexity because the upper tips are too small. However, zooming in helps, and in Fig.~\ref{vanvleck_2345b_s1_1.7_tips} we can see that the overall convex 4- and 5-bit curves have the inflection points and become concave near $|\rho|\approx1$. Plotting a family of the curves with one of the $\sigma$ fixed and the other one varied elicits the convexity alternation, as shown in Fig.~\ref{vanvleck_s1_0.6_s2_0.55_to_0.7}. With fixed $\sigma_x=0.6$ and $\sigma_y$ varying from 0.55 to 0.7 with the step 0.01, the upper end transformations are resemblant of a ``whip tip" in motion. The study of second derivative signs of the upper tips for all the 4-bit curves stored in our Van Vleck correction table over the 0.01-step grid of 301~x~301 $\sigma_x$ and $\sigma_y$ combinations showed that the convex ends constitute only small percentage of the majority of the concave curve ends. Fig.~\ref{where_curve_convex} visualizes a square fragment of the $(\sigma_x,\sigma_y)$ grid for $\sigma\approx0.4\ldots1.4$. The locations where the curves have convex upper tips are marked with black dots. One can see that all the diagonal elements, where $\sigma_x=\sigma_y$, are filled with the black dots. This means the Van Vleck correction functions for equal STDs of both input signals are always convex at their upper ends. The general distribution of convex tips, although symmetric with respect to the diagonal, looks otherwise quite irregular. 

\section{Conclusion}

In this work we have derived a generalized method for correction of the correlation computed on quantized signals. Our method is the direct expansion of the classical \citet{Vanvleck1966} correction formula~\eqref{vanvlec}. The method is derived for a general quantization schemes with an arbitrary number of levels and transition thresholds, as shown in Fig.~\ref{gen_qpattern}. A benefit of the method is its ability to correct the correlation of signals with different RMS. The method is based on the inversion of Eq.~\eqref{qkappa_as_integral}, which relates the quantized covariance (i.e. the correlator output) and the exact correlation. The equation includes the analog RMS of both input signals. We show that the standard deviations, computed for the quantized signals, cannot be used for their RMS because of the systematic errors (similar to the error in quantized correlation), and therefore require correction before being used in the correlation correction. This problem of estimation of the analog signal RMS from the quantized signal STD for general quantization patterns is solved using the inverse of Eq.~\eqref{qsig_general}. 

Since we are focusing on improvement of the imaging dynamic range of the Murchison Widefield Array (see, for example,  \citet{Lonsdale2009}), we provide versions of the general formulas restricted to the regular quantization patterns, used in the 4-bit MWA quantizer. The correlation correction is expressed as the inverse of Eq.~\eqref{qkappa_as_integral_mwa}, and the analog RMS is estimated as the inverse of Eq.~\eqref{sig_as_fun_qsig_regular}. 

We examined some interesting properties of the correction functions. In particular, unlike the classical Van Vleck function~\eqref{vanvlec_rho}, the curves of inverses of~\eqref{qkappa_as_integral} are convex for only some pairs of the signal RMS values in the vicinity of $\rho=1$. For the majority of signal RMS values the upper tips of the curves are concave. Fig.~\ref{where_curve_convex} helps build a general impression of this phenomenon. Note that for equal RMS values of the both correlated signals the correction functions are always convex at their $\rho=1$ ends (see the diagonal). 

The simulations demonstrated that for the 4-bit correlator the typical correlation error of $6 \pm 5\%$ after applying our correction method is reduced to a small value within $\pm0.1\%$. This, however, is only true as long as there are not too many outliers, while the typical error is within $\pm0.01 - 0.001\%$.  For these results please see Fig.~\ref{fig_means_errorbars_all_lev}. The latter is important in the context of the MWA solar observations with its typically highly correlated strong signals.

\section*{Acknowledgments}

This material is based upon work supported by the Air Force Office of Scientific Research (AFOSR) under Award No. FA9550-14-1-0192. Any opinions, findings, and conclusions or recommendations expressed in this publication are those of the author(s) and do not necessarily reflect the views of the AFOSR agency. We gratefully acknowledge the helpful discussions and input from Jim Moran, Randall Wayth, John Morgan, and Brian Crosse.


\bibliography{multilevel_Van_Vleck}

\begin{thebibliography}{26}
\providecommand{\natexlab}[1]{#1}
\expandafter\ifx\csname urlstyle\endcsname\relax
  \providecommand{\doi}[1]{doi:\discretionary{}{}{}#1}\else
  \providecommand{\doi}{doi:\discretionary{}{}{}\begingroup
  \urlstyle{rm}\Url}\fi

\bibitem[{\textit{{Bowman} et~al.}(2013)}]{Bowman2013}
{Bowman}, J.~D., et~al. (2013), {Science with the Murchison Widefield Array},
  \textit{Proc. Astron. Soc. Aust.}, \textit{30}, e031,
  \doi{10.1017/pas.2013.009}.

\bibitem[{\textit{Bracewell}(1986)}]{Bracewell1986}
Bracewell, R.~N. (1986), \textit{{The Fourier Transform and Its Applications}},
  2nd, revised ed., 474 pp., WCB/McGraw-Hill, Boston, Massachusetts.

\bibitem[{\textit{{Brown}}(1967)}]{Brown1967}
{Brown}, J.~L. (1967), {A Generalized Form of Price's Theorem and Its
  Converse}, \textit{IEEE Transactions on Information Theory}, \textit{13},
  27--30, \doi{10.1109/TIT.1967.1053965}.

\bibitem[{\textit{{Cooper}}(1970)}]{Cooper1970}
{Cooper}, B.~F.~C. (1970), {Correlators with two-bit quantization},
  \textit{Australian Journal of Physics}, \textit{23}, 521--527.

\bibitem[{\textit{{Greisen}}(2003)}]{Greisen2003}
{Greisen}, E.~W. (2003), {AIPS, the VLA, and the VLBA}, \textit{Information
  Handling in Astronomy - Historical Vistas}, \textit{285}, 109,
  \doi{10.1007/0-306-48080-8\_7}.

\bibitem[{\textit{{Hagen} and {Farley}}(1973)}]{Hagen1973}
{Hagen}, J.~B., and D.~T. {Farley} (1973), {Digital-correlation techniques in
  radio science.}, \textit{Radio Science}, \textit{8}, 775--784,
  \doi{10.1029/RS008i008p00775}.

\bibitem[{\textit{{Johnson} et~al.}(2013)\textit{{Johnson}, {Chou}, and
  {Gwinn}}}]{Johnson2013}
{Johnson}, M.~D., H.~H. {Chou}, and C.~R. {Gwinn} (2013), {Optimal Correlation
  Estimators for Quantized Signals}, \textit{Astrophysical Journal},
  \textit{765}, 135, \doi{10.1088/0004-637X/765/2/135}.

\bibitem[{\textit{{Kulkarni} and {Heiles}}(1980)}]{Kulkarni1980}
{Kulkarni}, S.~R., and C.~{Heiles} (1980), {How to obtain the true correlation
  from a 3-level digital correlator}, \textit{Astronomical Journal},
  \textit{85}, 1413--1420, \doi{10.1086/112815}.

\bibitem[{\textit{{Lonsdale} et~al.}(2009)}]{Lonsdale2009}
{Lonsdale}, C.~J., et~al. (2009), {The Murchison Widefield Array: Design
  Overview}, \textit{IEEE Proceedings}, \textit{97}, 1497--1506,
  \doi{10.1109/JPROC.2009.2017564}.

\bibitem[{\textit{{McMahon}}(1964)}]{McMahon1964}
{McMahon}, E.~L. (1964), {An Extension of Price's Theorem}, \textit{IEEE
  Transactions on Information Theory}, \textit{IT-10},
  \doi{10.1109/TIT.1964.1053656}.

\bibitem[{\textit{{McMullin} et~al.}(2007)\textit{{McMullin}, {Waters},
  {Schiebel}, {Young}, and {Golap}}}]{McMullin2007}
{McMullin}, J.~P., B.~{Waters}, D.~{Schiebel}, W.~{Young}, and K.~{Golap}
  (2007), {CASA Architecture and Applications}, in \textit{Astronomical Data
  Analysis Software and Systems XVI}, \textit{Astronomical Society of the
  Pacific Conference Series}, vol. 376, edited by R.~A. {Shaw}, F.~{Hill}, and
  D.~J. {Bell}, p. 127.

\bibitem[{\textit{{Oberoi} and {Benkevitch}}(2010)}]{OberoiBenkevitch2010}
{Oberoi}, D., and L.~{Benkevitch} (2010), {Remote Sensing of the Heliosphere
  with the Murchison Widefield Array}, \textit{Solar Physics}, \textit{265},
  293--307, \doi{10.1007/s11207-010-9580-x}.

\bibitem[{\textit{{Oberoi} et~al.}(2013)}]{Oberoi2013}
{Oberoi}, D., et~al. (2013), {Imaging the Sun with the Murchison Widefield
  Array}, in \textit{Astronomical Society of India Conference Series},
  \textit{Astronomical Society of India Conference Series}, vol.~10, pp.
  131--135.

\bibitem[{\textit{{Oberoi} et~al.}(2014)}]{Oberoi2014}
{Oberoi}, D., et~al. (2014), {Observing the Sun with the Murchison Widefield
  Array}, \textit{ArXiv e-prints}.

\bibitem[{\textit{{Ord} et~al.}(2015)}]{Ord2015}
{Ord}, S.~M., et~al. (2015), {The Murchison Widefield Array Correlator},
  \textit{Proc. Astron. Soc. Aust.}, \textit{32}, e006,
  \doi{10.1017/pasa.2015.5}.

\bibitem[{\textit{{Papoulis}}(1965)}]{Papoulis1965}
{Papoulis}, A. (1965), {Comment on 'An Extension of Price's Theorem' by
  McMahon}, \textit{IEEE Transactions on Information Theory}, \textit{IT-11},
  \doi{10.1109/TIT.1965.1053722}.

\bibitem[{\textit{{Prabu} et~al.}(2015)}]{Prabu2015}
{Prabu}, T., et~al. (2015), {A digital-receiver for the Murchison Widefield
  Array}, \textit{Experimental Astronomy}, \textit{39}, 73--93,
  \doi{10.1007/s10686-015-9444-3}.

\bibitem[{\textit{Press et~al.}(1992)\textit{Press, Teukolsky, Vetterling, and
  Flannery}}]{Numrecp1992}
Press, W.~H., S.~A. Teukolsky, W.~T. Vetterling, and B.~P. Flannery (Eds.)
  (1992), \textit{{Numerical Recipes in Fortran. The Art of Scientific
  Computing. Second Edition}}, 2nd ed., 963 pp., Cambridge Univ. Press, New
  York.

\bibitem[{\textit{{Price}}(1958)}]{Price1958}
{Price}, R. (1958), {A useful theorem for nonlinear devices having Gaussian
  inputs}, \textit{IRE Transactions on Information Theory}, \textit{42},
  RS3022, \doi{10.1109/TIT.1958.1057444}.

\bibitem[{\textit{{Sault} et~al.}(1995)\textit{{Sault}, {Teuben}, and
  {Wright}}}]{Sault1995}
{Sault}, R.~J., P.~J. {Teuben}, and M.~C.~H. {Wright} (1995), {A Retrospective
  View of MIRIAD}, in \textit{Astronomical Data Analysis Software and Systems
  IV}, \textit{Astronomical Society of the Pacific Conference Series}, vol.~77,
  edited by R.~A. {Shaw}, H.~E. {Payne}, and J.~J.~E. {Hayes}, p. 433.

\bibitem[{\textit{Thompson et~al.}(2001)\textit{Thompson, Moran, and
  Swensson}}]{TMS2001}
Thompson, A.~R., J.~M. Moran, and G.~W. Swensson, Jr. (2001),
  \textit{{Interferometry and Synthesis in Radio Astronomy. Second Edition}},
  2nd ed., 692 pp., John Wiley \& Sons, Inc, New York.

\bibitem[{\textit{{Thompson} et~al.}(2007)\textit{{Thompson}, {Emerson}, and
  {Schwab}}}]{Thomson2007}
{Thompson}, A.~R., D.~T. {Emerson}, and F.~R. {Schwab} (2007), {Convenient
  formulas for quantization efficiency}, \textit{Radio Science}, \textit{42},
  RS3022, \doi{10.1029/2006RS003585}.

\bibitem[{\textit{{Tingay} et~al.}(2013)}]{Tingay2013}
{Tingay}, S.~J., et~al. (2013), {The Murchison Widefield Array: The Square
  Kilometre Array Precursor at Low Radio Frequencies}, \textit{Proc. Astron.
  Soc. Aust.}, \textit{30}, e007, \doi{10.1017/pasa.2012.007}.

\bibitem[{\textit{{Tremblay} et~al.}(2015)}]{Tremblay2015}
{Tremblay}, S.~E., et~al. (2015), {The High Time and Frequency Resolution
  Capabilities of the Murchison Widefield Array}, \textit{pasa}, \textit{32},
  e005, \doi{10.1017/pasa.2015.6}.

\bibitem[{\textit{{Van Vleck} and {Middleton}}(1966)}]{Vanvleck1966}
{Van Vleck}, J.~H., and D.~{Middleton} (1966), {The Spectrum of Clipped Noise},
  \textit{Proceedings of the IEEE}, \textit{54}, 2--19,
  \doi{10.1109/PROC.1966.4567}.

\bibitem[{\textit{{Wayth} et~al.}(2009)\textit{{Wayth}, {Greenhill}, and
  {Briggs}}}]{Wayth2009}
{Wayth}, R.~B., L.~J. {Greenhill}, and F.~H. {Briggs} (2009), {A GPU-based
  Real-time Software Correlation System for the Murchison Widefield Array
  Prototype}, \textit{pasp}, \textit{121}, 857--865, \doi{10.1086/605334}.

\end{thebibliography}

\end{document}